\documentclass[aps,prd,nofootinbib,floatfix,showpacs,notitlepage,eqsecnum,twocolumn,10pt]{revtex4-2}

\usepackage{graphicx,wrapfig,slashed,bbold,bm}
\usepackage{amsmath,amssymb,epsfig,graphicx}
\usepackage[dvipsnames]{xcolor}
\usepackage{braket}
\usepackage{epstopdf}
\usepackage{multirow}
\usepackage{booktabs}
\usepackage{ragged2e}
\usepackage{mciteplus}
\usepackage[colorlinks=true,linkcolor=blue,urlcolor=blue,citecolor=blue]{hyperref}
\usepackage{cleveref}
\usepackage{simpler-wick}

\usepackage{chngcntr}
\counterwithout{equation}{section}

\usepackage{cancel}

\usepackage{makecell}
\usepackage{tikz}
\newcommand*\circled[1]{\tikz[baseline=(char.base)]{
            \node[shape=circle,draw,inner sep=1pt] (char) {#1};}}

\newcommand{\nc}{\newcommand}
\nc{\non}{\nonumber}
\nc{\hc}{\hbox {H.c.}}
\nc{\noi}{\noindent}
\nc{\barx}{\bar{x}}
\nc{\pbarn}{\;\hbox {pb}}
\nc{\fbarn}{\;\hbox {fb}}
\nc{\cone}{{\scriptsize \textcolor{orange}{\circled{1}}}}
\nc{\ctwo}{{\scriptsize \textcolor{orange}{\circled{2}}}}
\nc{\cthr}{{\scriptsize \textcolor{orange}{\circled{3}}}}

\nc{\hsp}{\hspace{0.5cm}}
\nc{\lsp}{\hspace{1cm}}
\nc{\Lsp}{\hspace{2cm}}
\nc{\LLsp}{\lsp\lsp}
\nc{\lra}{\longrightarrow}
\nc{\p}{\prime}
\nc{\sgn}{\text{sgn}}
\nc{\ph}{\varphi}
\nc{\op}{{\cal O}}
\nc{\tr}{\mathrm{tr}}
\nc{\eff}{\mathrm{eff}}
\nc{\sqM}{\sqrt{M}}
\nc{\NL}{\mathrm{NL}}
\nc{\moc}{\mathcal{M}} 
\nc{\rd}[1]{\textcolor{red}{#1}}
\nc{\bl}[1]{\textcolor{blue}{#1}}

\nc{\pbar}{\bar{\psi}}
\nc{\beq}{\begin{equation}}  \nc{\eeq}{\end{equation}}
\nc{\bea}{\begin{eqnarray}}  \nc{\eea}{\end{eqnarray}}
\nc{\baa}{\begin{array}}     \nc{\eaa}{\end{array}}
\nc{\bit}{\begin{itemize}}   \nc{\eit}{\end{itemize}}
\nc{\ben}{\begin{enumerate}} \nc{\een}{\end{enumerate}}
\nc{\bce}{\begin{center}}    \nc{\ece}{\end{center}}
\nc{\bpm}{\begin{pmatrix}}   \nc{\epm}{\end{pmatrix}}
\nc{\bvt}{\begin{verbatim}}  \nc{\evt}{\end{verbatim}}
\nc{\vp}[1]{\mathbf{p}_{#1}}
\nc{\vk}{{\bm k}_\perp}
\nc{\vkp}{{\bm k}'_\perp}
\nc{\vq}{{\bm q}_\perp}
\nc{\uu}{{\uparrow\uparrow}}
\nc{\ud}{{\uparrow\downarrow}}
\nc{\du}{{\downarrow\uparrow}}
\nc{\dd}{{\downarrow\downarrow}}

\def\lsim{\mathrel{\raise.3ex\hbox{$<$\kern-.75em\lower1ex\hbox{$\sim$}}}}
\def\gsim{\mathrel{\raise.3ex\hbox{$>$\kern-.75em\lower1ex\hbox{$\sim$}}}}

\def\udots{\mathinner{\mkern1mu\raise1pt\vbox{\kern7pt\hbox{.}}\mkern2mu\raise4pt\hbox{.}\mkern2mu\raise7pt\hbox{.}\mkern1mu}}

\def\la{\langle}
\def\ra{\rangle}

\def\vp{{\bf p}}

\def\lam{\lambda}

\def\be{\begin{equation}}
\def\ee{\end{equation}}
\def\bea{\begin{eqnarray}}
\def\eea{\end{eqnarray}}

\def\la{\langle}
\def\ra{\rangle}

\def\vp{{\bf p}}

\def\lam{\lambda}

\def\be{\begin{equation}}
\def\ee{\end{equation}}
\def\bea{\begin{eqnarray}}
\def\eea{\end{eqnarray}}

\allowdisplaybreaks

\begin{document}
\preprint{\begin{flushright}
Preprint number here\\  
\end{flushright}}

\author{Yongwoo Choi}
\email[ E-mail: ]{sunctchoi@gmail.com}
\affiliation{Department of Physics,
 Inha University, Incheon 22212, Republic of Korea}

\author{Hyeon-Dong Son}
\email[ E-mail: ]{hdson@inha.ac.kr}
\affiliation{Department of Physics,
 Inha University, Incheon 22212, Republic of Korea}
 
\author{Ho-Meoyng Choi}
\email[ E-mail: ]{homyoung@knu.ac.kr}
\affiliation{Department of Physics,
 Teachers College, Kyungpook National University, Daegu 41566, Republic of Korea}

\title{Gravitational form factors of the pion in the self-consistent light-front quark model} 

\begin{abstract}
We present a self-consistent light-front quark model (LFQM) analysis of the pion's gravitational form factors (GFFs), incorporating the
Bakamjian-Thomas (BT) construction consistently throughout the framework. By uniformly applying the BT formalism  to both hadronic matrix
elements and their associated Lorentz structures, we achieve a current-component-independent extraction of the pion GFFs $A_\pi(t)$ and $D_\pi(t)$,
thereby eliminating the light-front zero-mode ambiguities that typically hinder conventional LFQM approaches.
By tuning the model parameters, we identify an optimal set that successfully reproduces 
the decay constant and electromagnetic form factor of the pion, while yielding a $D$-term value $D_\pi(0) \approx -1$, consistent with predictions from chiral perturbation theory. 
The $D$-term emerges as a sensitive probe of the pion's internal dynamics, governing its mechanical radius---the largest among the charge,
mass, and mechanical radii. We further examine the pion's spatial structure via its associated two-dimensional light-front densities, including the
momentum density, transverse pressure, and shear stress, all of which satisfy the required normalization and von Laue stability conditions. 
Our results reveal a detailed mechanical landscape: a centrally peaked momentum density that decreases monotonically;
a repulsive pressure near the center (up to $x_\perp = 0.33$~fm) that transitions to attraction in the outer region; 
and a shear stress profile peaking at an intermediate distance ($x_\perp \approx  0.2$~fm).
\end{abstract}


\maketitle
\flushbottom
\section{Introduction}\label{Sec:I}
Hadronic matrix elements of the quantum chromodynamics (QCD) energy-momentum tensor (EMT)
provide fundamental insights into the intrinsic properties of hadrons,
including their mass, spin~\cite{Kobzarev:1962wt,Pagels:1966zza}, and internal structure.
 These properties emerge from the quark and gluon degrees of freedom
and are characterized by gravitational form factors (GFFs).
In particular, EMT matrix elements offer a unique perspective on the mechanical structure of hadrons through the $D$-term form factor,
which governs the pressure and shear distributions of quarks and gluons inside hadrons~\cite{Polyakov:2002yz}. 
These distributions, constrained by global and local stability 
conditions~\cite{Lorce:2018egm,Panteleeva:2020ejw,Kim:2020nug},  
determine how internal forces counteract one another to prevent a hadron from collapsing or expanding.
Comprehensive discussions of these aspects can be found in~\cite{Polyakov:2018zvc,Lorce:2025oot}.

Despite their fundamental importance, direct measurements of GFFs remain elusive. 
Instead, several indirect approaches have been developed to access information about the energy-momentum structure of hadrons.
One widely studied method involves the second Mellin moments of twist-2 vector generalized parton distributions (GPDs)~\cite{Muller:1994ses}, 
which encode both the transverse spatial structure and longitudinal momentum distributions of quarks and gluons inside hadrons.
The Mellin moments of GPDs yield fundamental form factors: the first moment 
corresponds to the electromagnetic (EM) form factor, which describes the charge distribution, while the
second moment gives the GFFs.

In the deeply virtual limit, the leading-order QCD factorization relates the amplitudes of
deeply virtual Compton scattering (DVCS) and deeply virtual meson production (DVMP) to GPDs, providing a valuable avenue for accessing GFFs. 
Recent experimental efforts have also explored alternative approaches to probe the proton's mechanical structure,
including the use of dispersion relation techniques applied to the measured Compton form factor in DVCS~\cite{Burkert:2018bqq}, and 
the extraction of gluonic EMT form factors from near-threshold photoproduction of the $J/\psi$ meson~\cite{Duran:2022xag,GlueX:2023pev,Guo:2023pqw}.

The pion, as a (pseudo) Nambu-Goldstone boson arising from spontaneously broken chiral symmetry, plays a fundamental role 
in low-energy QCD.
Understanding its internal mechanical structure--particularly in relation to the $D$-term--is of great interest.
Although experimental investigations into the pion's structure--especially its GPDs and GFFs--remain significantly 
more limited than those for the proton, there have been notable theoretical efforts to study the pion's GFFs.
The study in~\cite{Son:2014sna} examined the connection between the pion's mechanical stability and patterns of chiral symmetry breaking.
Progress has also been made~\cite{Kumano:2017lhr} in extracting pion GFFs from generalized distribution amplitudes (GDAs), 
using data from the Belle experiment~\cite{Belle:2015oin}. 
A recent theoretical proposal~\cite{Amrath:2008vx} suggests probing the pion's structure via virtual pion
exchange in future electron-ion collider (EIC) experiments~\cite{Chavez:2021koz,Hatta:2025ryj}. 
In parallel, lattice QCD~\cite{Brommel:2007zz,QCDSF:2007ifr,Shanahan:2018pib,Pefkou:2021fni,Hackett:2023nkr}, 
chiral perturbation theory (ChPT)~\cite{Donoghue:1991qv}, and various effective models~\cite{WB08,Freese_2019,Krutov21,ZXing,xu2023,YLi04,Zahed1,Zahed2,WB24}
have provided valuable theoretical insights into the pion's gravitational structure.

In the light-front (LF) formalism, GFFs are naturally expressed as overlap integrals of light-front wave functions (LFWFs), 
offering a direct microscopic interpretation of the hadronic EMT~\cite{BD2001}. 
This formulation enables a frame-independent evaluation of GFFs, which is a key advantage of the LF approach. 
However, extracting the $D$-term in light-front dynamics (LFD) presents a significant challenge, 
as it inevitably involves the so-called “bad” current components--such as the minus component, 
as well as certain combinations of perpendicular components in the EMT matrix elements--which are highly sensitive to LF zero modes. 
Addressing this issue is crucial for achieving a consistent and reliable formulation of the hadronic mechanical structure
within the LFD framework.

A self-consistent light-front quark model (LFQM) has been
developed~\cite{CJ14,CJ15,CJ17,Jafar1,Jafar2,Choi21,ChoiAdv,Choi:2024ptc},
ensuring that physical observables can be extracted in a current-component independent manner. 
This approach is rooted in the Bakamjian-Thomas (BT) construction~\cite{BT53,Poly10,CCKP},
which systematically incorporates interactions while preserving relativistic consistency. 
In this framework, meson states are constructed from noninteracting, on-mass-shell quark and antiquark 
states, with interactions introduced via the total mass operator $M:= M_0 + V_{q{\bar q}}$.
The on-mass shell condition ensures four-momentum conservation
at the meson-quark vertex, $p=p_{q} + p_{{\bar q}}$, 
where $p$ and $p_{q(\bar q)}$ denote the meson and constituent quark (antiquark) momenta, respectively.
In particular, conservation of LF energy at the meson--quark vertex, expressed as
$p^- = p^-_q + p^-_{\bar q}$,
underscores the need to identify the meson mass with  the invariant mass $M_0$ constructed from the quark and antiquark momenta. 
This ensures the consistency condition
$\frac{M_0^2 + {\bm p}^2_\perp}{p^+} 
= \frac{m_q^2 + {\bm p}^2_{q\perp}}{p^+_q}
+ \frac{m_{\bar q}^2 + {\bm p}^2_{{\bar q}\perp}}{p^+_{\bar q}}$
is satisfied in the calculation of the meson--quark vertex.
This equivalence justifies the substitution $M\to M_0$, which is key feature of the BT-based LFQM.
This framework has been extensively applied to the mass spectra of pseudoscalar and vector mesons, 
including both ground and radially excited states~\cite{CJ97,CJ99a,CJLR15,ACJO22}, 
demonstrating its effectiveness across a wide range of mesonic systems.

While the BT construction is widely used in relativistic quark models to ensure 
Poincar\'{e} invariance, its implementation in LFQM studies has often been incomplete.
For example, when evaluating the hadronic matrix element for a local current $\mathcal{O}^\mu$,
\begin{align}\label{eq:1}
\la p'| \mathcal{O}^{\mu} |p\ra =\Gamma^\mu {\cal F}, 
\end{align}
where ${\cal F}$ denotes a physical observable (e.g., decay constant, form factor, etc.) 
and $\Gamma^\mu$ encodes the Lorentz structure, 
the conventional LFQM~\cite{SLF2,SLF3,CCC97} evaluate ${\cal F}$ as
\begin{align}\label{eq:2}
{\cal F}=\frac{1}{{\Gamma}^\mu}\bra{p'}\mathcal{O}^\mu \ket{p}_{\rm BT}.
\end{align}
Here, the subscript “BT” indicates that the matrix element is evaluated using the BT construction, 
i.e., with the physical meson mass $M$ replaced by the invariant mass $M_0$ in the internal kinematics.
Physically,  the meson mass $M$ characterizes the bound state, while  the invariant mass $M_0$ 
reflects internal kinematics and enters LF wave functions and matrix elements.
Inconsistent use of $M$ and $M_0$--as seen in conventional LFQM implementations--can lead to kinematic 
mismatches and spurious LF zero mode contributions. 
These typically arise when the BT construction is applied only to the matrix element but not 
to the full Lorentz structure~\cite{SLF2,SLF3,CCC97}, especially when “bad” current components are used.

To resolve this, 
we proposed~\cite{CJ14,CJ15,CJ17,Jafar1,Jafar2,Choi21,ChoiAdv,Choi:2024ptc} a self-consistent LFQM 
that applies the BT construction uniformly to both hadronic matrix elements and Lorentz structures. 
This unified treatment ensures that the observable ${\cal F}$ remains independent of the current component. 
Specifically, instead of computing the observable as in Eq.~\eqref{eq:2}, we implement
\begin{align}\label{eq:3}
{\cal F}=\bra{p'}\frac{\mathcal{O}^\mu }{{\Gamma}^\mu}\ket{p}_{\rm BT}, 
\end{align}
where the Lorentz structure $\Gamma^\mu$ is consistently incorporated into the integral.
This formulation guarantees that the observable 
${\cal F}$ is independent of the current component used and free of zero-mode contamination.

As a consequence of this consistent treatment, the most general Lorentz structures for hadronic matrix elements 
must accomodate transitions between states with different initial and final masses 
($M\neq M'$)--even in diagonal matrix elements. This is because, within the BT framework, the internal invariant
masses differ ($M_0\neq M'_0$), leading to non-vanishing contributions from terms involving $M'^2-M^2$ even when $M = M'$. 

Our formulation has been systematically tested and validated across a wide class of observables:
(1) decay constants and distribution amplitudes (DAs), including higher-twist contributions, for pseudoscalar and vector mesons~\cite{CJ14,CJ15,CJ17,Jafar1,Jafar2}, 
(2) weak transition form factors between pseudoscalar mesons~\cite{Choi21,ChoiAdv}, and
(3) the pion's EM form factor and transverse momentum-dependent distributions (TMDs)~\cite{Choi:2024ptc}.
Furthermore, our self-consistent LFQM can be systematically derived from the covariant Bethe-Salpeter (BS) model~\cite{CJ14,CJ15,CJ17,Jafar1,Jafar2,Choi21,ChoiAdv}
using a type-II matching condition (e.g., Eq. (49) in~\cite{CJ14}) that ensures full compatibility with the BT framework.

In this work, we extend our self-consistent LFQM~\cite{CJ14,CJ15,CJ17,Jafar1,Jafar2,Choi21,ChoiAdv,Choi:2024ptc} to compute the pion's GFFs and 
demonstrate that they can be extracted in a current-component-independent manner.
Importantly, the extracted pion GFFs,  $A_\pi(t)$ and $D_\pi(t)$, satisfy essential physical 
constraints--such as $A_\pi(0) = 1$ and $\lim_{t \to 0} tD_\pi(t) = 0$--and are expressed in terms of physical momenta, 
making them suitable for extrapolation to physically 
meaningful quantities such as pressure and shear distributions.

The paper is organized as follows: 
Section~\ref{Sec:II} introduces our self-consistent LFQM, fully incorporating the BT construction.
Section~\ref{Sec:III} presents the methodology for computing the pion's GFFs, $A_\pi(t)$ and $D_\pi(t)$, 
including the transition process $\pi(p)\to\pi(p')$, with the physical masses $M$ and $M'$ replaced by
the invariant masses
$M_0$ and $M'_0$. Both GFFs are shown to be extractable in a current-component–independent manner.
Section~\ref{Sec:IV} presents our numerical results for the pion's EM form factor and GFFs,
and investigate possible LF zero-mode contributions to the GFFs that may arise in the conventional LFQM.
We also analyze the spatial structure of the pion through transverse densities--such as LF momentum density, 
pressure, and shear stress--providing a mechanical interpretation.
Finally, Sec.~\ref{Sec:V} summarizes our findings.

\section{Light-Front Quark Model}\label{Sec:II}
A key feature of the LFQM~\cite{SLF2,SLF3,CP05,CJ97,CJ99a,CJ07} for $q{\bar q}$ bound-state mesons is the simplified Fock-state expansion,
which considers only the constituent quark and antiquark. The Fock state is treated within a non-interacting $q{\bar q}$ framework, 
with interactions incorporated into the mass operator as $M: = M_0 + V_{q\bar{q}}$, 
ensuring compatibility with the relevant symmetry group structures and commutation relations. 
These interactions are encapsulated in the LFWF $\Psi_{\lambda_1\lambda_{2}}^{JJ_z}(\bm{p}_1,\bm{p}_{2})$, 
which serves as an eigenfunction of the mass operator and
describes the internal structure of the hadron. Here, $\lambda_{1(2)}$ denotes the helicities 
of the constituent quark (antiquark), and $(J,J_z)$ represents the total angular momentum of the hadron.

The four-momentum $p$ of a meson in LF coordinates is expressed as $p=(p^+, p^-,  \bm{p}_\perp)$, where
$p^{+}=p^0 + p^3$ represents the longitudinal LF momentum, $p^-=p^0-p^3$ corresponds to the LF energy, and
$\bm{p}_\perp=(p^1, p^2)$ denotes the transverse momentum components. 
We adopt the Minkowski metric convention, in which the inner product of two four-vectors is defined as
$a\cdot b=\frac{1}{2} (a^+b^- + a^-b^+) - \bm{a}_\perp\cdot \bm{b}_\perp$.
The pion state $\ket{\pi (p, J=0, J_z=0)}$ can be expanded in terms of its quark-antiquark Fock state as
\begin{align}\label{eq:4}
\ket{\pi(p)}
&= \int \left[ d^3\bm{p}_1 \right] \left[ d^3\bm{p}_{2} \right]  2(2\pi)^3 \delta^{(3)} \left(\bm{p}-\bm{p}_1-\bm{p}_{2} \right) 
\nonumber\\ & \times \mbox{} 
\sum_{\lambda_1,\lambda_{2}} \Psi^{JJ_z}_{\lambda_1 \lambda_{2}}(\bm{p}_1,\bm{p}_{2}) 
\ket{q(p_1,\lambda_1) \bar{q}(p_{2},\lambda_{2}) }.
\quad
\end{align}
Here, $\bm{p}=(p^{+},\bm{p}_{\perp})$ denotes the LF three-momentum and the integration
measure is given by $\left[ d^3\bm{p} \right]=\frac{{\rm d}p^+ {\rm d}^2\bm{p}_{\perp}}{2(2\pi)^3}$.
The two-particle state $\ket{q(p_1,\lambda_1) \bar{q}(p_2,\lambda_2)}$ is understood as being created 
by the action of the quark and antiquark creation operators on the vacuum:
\bea\label{eq:creation}
\ket{q(p_1,\lambda_1) \bar{q}(p_2,\lambda_2)} = a^\dagger_{\lambda_1}(p_1)\, b^\dagger_{\lambda_2}(p_2)\ket{0},
\eea
where $a^\dagger$ and $b^\dagger$ respectively denote the quark and antiquark creation operators,
which satisfy standard fermionic anticommutation relations:
\bea\label{eq:anticom}
\{ a_{\lambda'}(p'),a^\dagger_\lambda(p)\}&=&\{ b_{\lambda'}(p'),b^\dagger_\lambda(p)\}\nonumber\\
&=& 2 (2\pi)^3 \delta^{(3)}({\bm p'}-{\bm p})\delta_{\lambda'\lambda}.
\eea
The LF relative momentum variables $(x, \bm{k}_\perp)$ are defined as $x_i=p^+_i/p^+$ and $\bm{k}_{i\perp}=\bm{p}_{i\perp}-x_i\bm{p}_\perp$,
satisfying the momentum conservation conditions $\sum_i x_i=1$ and $\sum_i \bm{k}_{i\perp}=0$. 
In this work, we define $x_1\equiv x$ and $\bm{k}_{1\perp}\equiv  \bm{k}_{\perp}$.
 
We normalize the pion state as 
\bea\label{eq:statenorm}
\Braket{\pi(p')|\pi(p)}=2 (2\pi)^3 p^+  \delta^{(3)}({\bm p'}-{\bm p})\delta_{J'J}\delta_{J'_z J_z},
\eea
so that the LFWF satisfies the normalization condition:
\be\label{eq:Psinorm}
1 = \int^1_0 dx\int \frac{d^2\bm{k}_\perp}{2(2\pi)^3} \sum_{\lambda_1\lambda_2}
\Psi^{\dagger J'J'_z}_{\lam_1{\lam_2}}(x,{\bm k}_{\perp})\Psi^{JJ_z}_{\lam_1{\lam_2}}(x,{\bm k}_{\perp}).
\ee
The momentum space LFWF of the pion $(J=J_z=0)$ expressed in terms of $(x, {\bm k}_\perp)$, can be written as
\begin{align}\label{eq:6}
\Psi^{00}_{\lam_1{\lam_2}}(x,{\bm k}_{\perp})
=\phi(x,{\bm k}_{\perp}){\cal R}^{00}_{\lam_1{\lam_2}}(x,{\bm k}_{\perp}),
\end{align}
where ${\cal R}^{00}_{\lam_1{\lam_2}}$ denotes the spin-orbit wave function, obtained via the Melosh transformation~\cite{Melosh}.
This transformation is interaction-independent and relates the conventional instant-form spin-orbit structure (labeled by $J^{PC}$)
to its LF counterpart. 

In practice, it is convenient to express ${\cal R}^{00}_{\lam_1{\lam_2}}$ in a covariant form~\cite{SLF2,SLF3,CCC97}:
\be\label{eq:SO1}
{\cal R}^{00}_{\lam_1{\lam_2}}=\frac{\bar{u}_{\lam_1}(p_1)\gamma_5 v_{\lam_{2}}( p_{2})}{\sqrt{2}M_0},
\ee
or alternatively, in its explicit matrix form:
\begin{align}\label{eq:7}
{\cal R}^{00}_{\lam_1{\lam_{2}}}
=\frac{1}{\sqrt{2 x(1-x)}M_{0}}\left(
\begin{array}{cc}
        -k^{(-)}  & m  \\ 
       -m  & -k^{(+)} 
      \end{array}
    \right),\;
\end{align}
where $k^{(\pm)}=k_x \pm i k_y$ and 
\begin{align}\label{eq:5} 
M_{0}^2 =\frac{ {\bm k}_\perp^{2} + m^2}{x} + \frac{ {\bm k}_\perp^{2} + m^2}{1-x},
\end{align}
is the invariant mass squared of the quark-antiquark system.

In our model, the pion is treated as an isospin-symmetric $q{\bar q}$ bound state, where
the constituent quark and antiquark masses are taken to be equal ($m=m_1=m_2$) 
in accordance with SU(2) flavor symmetry.
We also note that the spin-orbit wave function ${\cal R}^{00}_{\lam_1{\lam_2}}$  satisfies the unitarity condition: 
$\sum_{\lambda's}{\cal R}^{\dagger} {\cal R}=1$.

The expression in Eq.~\eqref{eq:5} follows directly from the on-mass-shell condition for the constituent quark and antiquark,
together with the conservation of LF energy at the meson-quark vertex, i.e.,
$p^-=p^-_1 + p^-_2$. 
The relation between the longitudinal momentum fraction $x$ and the $z$-component $k_z$ of the quark's three-momentum
is given by~\cite{SLF2,SLF3} $k_{z}=\left(x-1/2\right)M_{0}$.
Thus, the Jacobian for the variable transformation
$(x,{\bm k}_\perp)\to ({\bm k}_\perp,k_z)$ is
$(\partial k_z/\partial x)=M_0/4 x (1-x)$.

The interaction between the quark and antiquark is incorporated into the mass operator~\cite{BT53,Poly10,CCKP},
enabling the computation of the meson's mass eigenvalue.
Within the LFQM framework, the radial wave function $\phi(x,{\bm k}_\perp)$ is treated as a variational trial function,
optimized for a QCD-motivated effective Hamiltonian. 
Further discussion of meson mass spectrum analyses can be found in Refs.~\cite{CJ97,CJ99a,CJ09,CJLR15,ACJO22,Nisha}.

For the $1S$ pion state, we adopt a Gaussian wave function as a trial wave function:
\begin{align}\label{eq:8}
\phi(x,{\bm k}_{\perp})=
\frac{4\pi^{3/4}}{\beta^{3/2}} \sqrt{\frac{\partial k_z}{\partial x}} {\rm exp}\left(-\frac{\vec{k}^2}{2\beta^2}\right),
\end{align}
where $\vec{k}^2= \bm{k}^2_\perp + k^2_z = M^2_0/4 -m^2$, and $\beta$ is a variational parameter determined 
from meson mass spectrum analyses~\cite{CJ97,CJ99a,CJ09}.
The state normalization given in Eq.~\eqref{eq:statenorm}, together with the unitarity of the spin-orbit wave function,
guarantees that  
the radial wave function satisfies the following normalization condition:
\begin{align}\label{eq:9}
\int^1_0 dx \int \frac{d^2{\bm k}_\perp}{2(2\pi)^3}~
|\phi(x,{\bm k}_{\perp})|^2=1.
\end{align}

\section{Gravitational form factors of the pion}\label{Sec:III}
We analyze the transition process in which a pion with momentum $p$ and physical mass $M$
transitions to a final state with momentum $p'$ and mass $M'$.
To describe the quark EMT operator, we adopt the following form:
\begin{align}\label{eq:10}
\hat T_q^{\mu\nu}(x) = \frac{1}{4}\bar \psi_q(x) \left(
i \gamma^\mu \overleftrightarrow{\partial}{}^\nu
+ i \gamma^\nu  \overleftrightarrow{\partial}{}^\mu
\right) \psi_q(x).
\end{align}
where $\overleftrightarrow{\partial}{}^\mu=\overrightarrow{\partial}{}^\mu-\overleftarrow{\partial}{}^\mu$.
The matrix element of the quark EMT operator between the initial and final pion states, $|\pi(p)\rangle$ and  $|\pi(p')\rangle$,
can be decomposed in terms of the quark GFFs, $A_q(t)$ and $D_q(t)$ as
\begin{align}\label{eq:11}
\braket{T^{\mu\nu}_q}&=\langle \pi(p') | \hat{T}^{\mu\nu}_q(0) | \pi(p) \rangle \cr
&~= {\Gamma}_A^{\mu\nu} A_q(t) + {\Gamma}_B^{\mu\nu} D_q(t) + 2m_\pi^2\bar c_q (t) g^{\mu\nu}.
\end{align}
Here, $A_q(t)$ describes the quark contribution to the energy and momentum densities, 
while $D_q(t)$ characterizes the quark contribution to the internal pressure and shear stress distributions. 
The form factor $\bar c_q(t)$ arises due to the non-conservation of the quark EMT for individual quark flavors
and contributes to the decomposition of the hadron's mass and mechanical structure. 

Under the assumption of SU(2) flavor symmetry for the pion ($m_u=m_d$), where both valence quarks
contribute equally to total form factors, the pion GFFs are given by
\begin{align}\label{eq:12}
A_\pi(t) = 2 A_q(t), \quad D_\pi(t) = 2D_q(t).
\end{align}
Since the total EMT is conserved, the form factor $\bar c_\pi(t)=2 \bar c_q(t)$ vanishes 
in our constituent quark model.

The Lorentz structures ${\Gamma}_A^{\mu\nu}$ and ${\Gamma}_D^{\mu\nu}$ are given by
\begin{align}\label{eq:13}
{\Gamma}_A^{\mu\nu} &= 2 \left[ P^\mu P^\nu - \frac{P\cdot \Delta}{\Delta^{2}}\left( \Delta^\mu P^\nu + \Delta^\nu P^\mu - P\cdot \Delta~ g^{\mu\nu}\right) \right],\nonumber\\
{\Gamma}_D^{\mu\nu} &= \frac{1}{2} \left(\Delta^\mu \Delta^\nu - g^{\mu\nu} \Delta^{2}\right),
\end{align}
where $P=\frac{p+p'}{2},~ \Delta = p'-p,~ P\cdot \Delta=\frac{M'^2-M^2}{2}$, and $\Delta^2=t<0$. 
In this work, we adopt $\Delta^+=0$ frame with ${\bm p}_\perp={\bf 0}_\perp$. 
In this frame, the four-momenta of the initial and final pion are given by
\begin{align}\label{eq:14}
    p &= \left(p^+,\frac{M^2}{p^+}, {\bm 0}_\perp \right), \nonumber\\
    p' &= \left(p^+, \frac{M'^2+ {\bm \Delta}_\perp^2}{p^+}, {\bm \Delta}_\perp \right).
\end{align}

We note that in the conventional LFQM~\cite{SLF2,SLF3,CCC97}, the Lorentz structure  
${\Gamma}_A^{\mu\nu} = 2 P^\mu P^\nu$ is typically used, and
the BT construction is applied only to the matrix elements $\braket{T^{\mu\nu}_q}$, not to the Lorentz structure itself.

In contrast, our BT-based formulation consistently incorporates the internal dynamics through the invariant masses 
$M_0$ and $M'_0$ of the initial and final states, which generally differ due to the momentum transfer $\Delta$, even when the external physical masses are equal ($M = M'$). 
This necessitates employing the full Lorentz structure given in Eq.~\eqref{eq:13}, 
including the term proportional to $P\cdot \Delta \propto (M'^2 - M^2)$, which under our BT construction consistently translates to 
$P\cdot \Delta \propto (M'^2_0 - M^2_0)$ 
in the matrix elements and their Lorentz decomposition.
This term plays a crucial role in ensuring the current conservation, $\partial_\mu \braket{T^{\mu\nu}_q}=0$,
and we have explicitly verified through numerical analysis that our matrix elements satisfy this condition,
thereby confirming that our normalization and mass conventions are fully consistent with EMT conservation.

The LF on-shell momenta $p_{1(2)}$ of the incoming quark (antiquark) and $p'_{1(2)}$ of the outgoing quark (antiquark)
for the transition $\ket{p(q{\bar q})}\to \ket{p' (q' {\bar q'})}$ are given by
\begin{align}\label{eq:15}
& p^+_i = p^{\prime +}_i =x_i p^+, 
\nonumber\\
& {\bm p}_{i\perp} = x_i {\bm p}_\perp + {\bm k}_{i\perp},~\; 
{\bm p}'_{i\perp} = x_i {\bm p}'_\perp + {\bm k}'_{i\perp},\; 
\end{align}
where $x_1=x$ and ${\bm k}_{1\perp}={\bm k}_\perp$.
Since the spectator antiquark satisfies $p^+_{2}=p'^+_{2}$ and ${\bm p}_{2\perp}={\bm p}'_{2\perp}$,
it follows that ${\bm k}'_\perp = {\bm k}_\perp + (1-x) {\bm \Delta}_\perp$.

In the LFQM framework, which adopts a noninteracting $q{\bar q}$ representation consistent with the BT construction,
the one-loop contribution to the matrix elements $\braket{T^{\mu\nu}_q}$ is obtained 
by convoluting the initial and final state LFWFs~\cite{Choi:2024ptc}:
\begin{align}\label{eq:16}
\braket{T^{\mu\nu}_q} = 
\int [d^3{\bm k}]\
\phi'(x,{\bm k}^\prime_\perp)  \phi(x,{\bm k}_\perp) S^{\mu\nu}_q,
\end{align}
where $S^{\mu\nu}_q$ represents the  spin trace for the tensor current, defined as 
\begin{align}\label{eq:17}
S^{\mu\nu}_q &=
 \mathcal{R}^\dagger_{\lambda'_1\lambda_2} 
\frac{\bar u_{\lambda'_1}(p'_1)}{\sqrt{p^{\prime +}_1}} 
    \bigg[ \frac{1}{4}\left(p_s^{\mu} \gamma^{\nu} + p_s^{\nu} \gamma^{\mu} \right) \bigg]
    \frac{ u_{\lambda_1}(p_1)}{\sqrt{p_1^+}}
\mathcal{R}_{\lambda_1\lambda_2}
\nonumber\\
&\equiv \mathcal{R}^\dagger_{\lambda'_1\lambda_2} 
U^{\mu\nu}_{\lambda_1\to\lambda'_1}(p_1\to p'_1)
\mathcal{R}_{\lambda_1\lambda_2},
\end{align}
with $p_s= p_1 + p'_1$, and the sum over quark helicities is implicit.

Applying the Dirac matrix elements for helicity spinors~\cite{BD80,Choi:2024ptc} to the tensor current,
we summarize $U^{\mu\nu}_{\lambda_1\to \lambda'_1}(p_1\to p'_1)$ in Table~\ref{tab:1}.
Additionally, Table~\ref{tab:2} presents the results for $S^{\mu\nu}$, $\Gamma_A^{\mu\nu}$, and $\Gamma_D^{\mu\nu}$ 
for all components except $(\mu\nu)=(--)$\footnote{ While single-minus current components such as $(+-)$ and $(-i)$
yield well-defined results, the double-minus component $(--)$ does not, reinforcing the need to avoid its use in the light-front GFF computation.}. 
Notably, for the transverse current components $(\mu\nu)=(ij)$ with $i\neq j$, the following relations hold:
\begin{align}\label{eq:18}
{\Gamma}_D^{ij}{\Delta}^i{\Delta}^j=0,
\;\;
{\Gamma}_A^{ij}{\Delta}^i{\Delta}^j=\frac{1}{2}(M^2_- + {\bm \Delta}^2_\perp)^2,
\end{align}
where  $M_-^{2}={M'}_{0}^{2}-M_{0}^{2}$. This result, together with the entries in Table~\ref{tab:2},
indicates that the form factor $A_q(t)$ can be independently extracted from three distinct
combinations of current components: $(\mu\nu)=(++), (+ i)$, $(ij)$. 
In contrast, the form factor $D_q(t)$ can be extracted from three other combinations: $(\mu\nu)=(+-), (- i)$, or
$(ii)$\footnote{Here, $(ii)$ implies the Einstein summation over $i=x,y$.},
in conjunction with the form factor $A_q(t)$.

\begin{table*}[hbt]
    \caption{\label{tab:1}%
    Matrix elements of $U^{\mu\nu}_{\lambda_1\to \lambda'_1}(p_1\to p'_1)$, as defined in Eq.\eqref{eq:17}. Here,
     $p_{(\pm)}=p^1\pm i p^2$ and $\delta_{(\pm)}^i=\delta^{i1}\pm i\delta^{i2}$. 
     Since \( p^+_1 = p^{\prime +}_1 \) and $\delta p_{(\pm)}=(p_{1(\pm)} - p'_{1(\pm)})$  hold in this framework, the results are expressed accordingly.}
    \begin{tabular}{ccc} 
    \hline \hline \\[-1ex]
    Matrix & \multicolumn{2}{c}{Helicity $\lambda_1 \to \lambda'_1$} \\ 
    elements  &  &  \\
    $U^{\mu\nu}$ & 
     $ \left\{ \begin{array}{c}
    \uparrow \;\to\; \uparrow \\
    \downarrow \;\to\;  \downarrow
   \end{array}  \right\} $ 
    & $ \left\{ \begin{array}{c}
    \uparrow \;\to\; \downarrow \\
    \downarrow \;\to\;  \uparrow
   \end{array}  \right\} $  \\[3.5ex]
    \hline \\
    $U^{++} $ & $2p_1^+$ & 0 \\[3.5ex]
    $U^{+-}$ & $\frac{1}{2}
    \left(p_1 + p'_1\right)^- + \frac{1}{p_1^+}\left(m^2+ p_{1(\pm)} p'_{1(\mp)} \right)$ & 
    $ \pm \frac{m}{p_1^+ }\delta p_{(\pm)} $\\[3.5ex]
    $U^{+i}$ &
        $\frac{1}{2}\left( p_{1}+ p'_{1}\right)^{i} + \frac{1}{2}
        \left( p_{1(\pm)} \delta^i_{(\mp)} + p'_{1(\mp)} \delta^i_{(\pm)} \right)$ & 0
        \\[3.5ex]
    $U^{-i}$ &
    $\frac{( p_{1}+ p'_{1})^i}{2(p_1^+)^{2}} \left(  m^2+ p_{1(\pm)} p'_{1(\mp)} \right)
     + 
    \frac{(p_{1} +p'_{1})^-}{4p_1^+} \left(  p_{1(\pm)} \delta^i_{(\mp)} + p'_{1(\mp)} \delta^i_{(\pm)} \right)
    $ & 
    $\pm\frac{ m( p_{1}+ p'_{1})^i}{2(p_1^+)^{2}} \delta p_{(\pm)}  $
    \\[3.5ex]
    $U^{ij}$ &~~~~ 
    $\frac{( p_{1}+ p'_{1})^i}{4 p_1^+}
    \left( p_{1(\pm)} \delta^j_{(\mp)} + p'_{1(\mp)} \delta^j_{(\pm)}  \right)
    +\frac{( p_{1}+ p'_{1})^j}{4 p_1^+}
    \left( p_{1(\pm)} \delta^i_{(\mp)} + p'_{1(\mp)} \delta^i_{(\pm)}  \right) $  ~~~~& 0
    \\[3.5ex]
      \hline\hline
    \end{tabular}
    \end{table*}
    
\begin{table*}
    \caption{\label{tab:2}%
    Spin traces for the tensor current $S^{\mu\nu}_q$, and Lorentz structures $\Gamma_A^{\mu\nu}$ and $\Gamma_D^{\mu\nu}$. 
    Note that we set the kinematics as $p_{1}^{+}={p'}_{1}^{+}$, and define $S_c = (m^2+ \vk \cdot \vkp)(\vk^2 + \vk\cdot\bm{\Delta}_{\perp} + m^2) + (1-x)(\vk\times \bm{\Delta}_{\perp})^2 + (1-x) m^2 \bm{\Delta}_{\perp}^2$, $M^2_{\pm}= {M}_{0}^{\prime 2} \pm M_0^2$.} 
    \begin{tabular}{c@{\hskip 0.5cm}c@{\hskip 0.5cm}c@{\hskip 0.5cm}c}
        \hline\hline \\[-1.5ex]
    $(\mu\nu)$ & $S^{\mu\nu}_q$ & $\Gamma_A^{\mu\nu}$ & $\Gamma_D^{\mu\nu}$ \\[1.5ex]
    \hline \\[-1.5ex]
    $(++)$ & $\frac{ 2p_1^+(m^2+ \vk \cdot \vkp)}{\sqrt{m^2 + \vk^2}\sqrt{m^2+\vk'^2}}$  & $2 (P^+)^2$ & 0 \\[3.5ex]
    $(+-)$ & $\frac{p^+_1 (p_1+p'_1)^-(m^2 + \vk\cdot\vk') + 2 S_c}{2 p^+_1 \sqrt{m^2 + \vk^2}\sqrt{m^2+\vk'^2}}$ 
    & $M^2_+  + M^2_- + \bm{\Delta}_{\perp}^2$ & $\bm{\Delta}_{\perp}^2$ \\[3.5ex]
     $(+i)$ & $ \frac{(2 k^i + \Delta^i)(m^2+ \vk \cdot \vkp)}{\sqrt{m^2 + \vk^2}\sqrt{m^2+\vk'^2}}$ 
     & $P^+ \Delta^i  \left(1+ \frac{M^2_-}{\bm{\Delta}_{\perp}^2}\right)$  & 0 \\[3.5ex]
     $(-i)$ & $\frac{(2k^i + \Delta^i) \left[\frac{1}{2}p^+_1(p_1+p'_1)^- (m^2+ \vk \cdot \vkp) + S_c\right]}
     {2(p^+_1)^{2}\sqrt{m^2 + \vk^2}\sqrt{m^2+\vk'^2}}$ 
     & $\frac{\Delta^i (\bm{\Delta}_{\perp}^2 +M^2_-)(\bm{\Delta}_{\perp}^2 + M^2_+ +M^2_-) }{2 \bm{\Delta}_{\perp}^2 P^+}$ & 
    $ \frac{\Delta^i (\bm{\Delta}_{\perp}^2+M^2_- )}{2P^+}$
     \\[3.5ex]
     $(ij)$ & $\frac{(2 k^i + \Delta^i)(2 k^j +\Delta^j)(m^2+ \vk \cdot \vkp)}{2 p^+_1\sqrt{m^2 + \vk^2}\sqrt{m^2+\vk'^2}}$ & $\frac{1}{2}\left[ \Delta^i \Delta^j\left(1 + \frac{2M^2_-}{\bm{\Delta}_{\perp}^2}\right) -g^{ij}\frac{M^4_-}{\bm{\Delta}_{\perp}^2}\right] $ &
    $\frac{1}{2}(\Delta^i \Delta^j + g^{ij}\bm{\Delta}_{\perp}^{2})$
      \\[3.5ex]
    \hline\hline
    \end{tabular}
\end{table*}

We define the generic form of the matrix element for the tensor current, incorporating the Lorentz factors 
in our LFQM consistent with the BT construction, as
\be\label{eq:19}
\Braket{\frac{T^{\mu\nu}_q}{{\Gamma}^{\mu\nu}_{A(D)}}}_{\rm BT} \equiv
\int [d^3{\bm k}]
\phi'(x, {\bm k}^\prime_\perp)  \phi(x, {\bm k}_\perp)  
\frac{S^{\mu\nu}_q}{{\Gamma}^{\mu\nu}_{A(D)}}.
\ee
In this expression, all occurrences of the meson masses $M^{(\prime)}$ in $\Gamma^{\mu\nu}_{A(D)}$
are consistently replaced with their corresponding invariant masses $M^{(\prime)}_0$. 
Specifically, $M'_0 \equiv M_0({\bm k}_\perp\rightarrow{\bm k}^{\prime}_\perp)$ 
denotes the invariant mass associated with the final-state pion.
Notably, the terms proportional to $P \cdot \Delta$ in $\Gamma^{\mu\nu}_{A}$ 
yield nonvanishing contributions due to the difference between the invariant masses (i.e., $M_0\neq M'_0$).

Using the definition in Eq.~\eqref{eq:19}, we compute
the form factor $A_q(t)$ from three different components, $(\mu\nu)=[(++), (+ i), (ij)]$, as follows:
\begin{align}\label{eq:20}
A^{(++)}_q(t) &= \Braket{\frac{T^{++}_q}{{\Gamma}_A^{++}}}_{\rm BT},
\nonumber\\
A^{(+i)}_q(t) &= \Braket{\frac{T^{+i}_q{{\Delta}^i}}{{\Gamma}_A^{+i} {{\Delta}^i}}}_{\rm BT},
\nonumber\\
A^{(ij)}_q(t)&= \Braket{\frac{T^{ij}_q{{\Delta}^i}{{\Delta}^j}}{{\Gamma}_A^{ij} {{\Delta}^i}{{\Delta}^j}}}_{\rm BT}.
\end{align}
We find that all three expressions yield identical results, confirming the internal
consistency of our LFQM formulation: $A_q(t)=A^{(++)}_q(t)=A^{(+i)}_q(t)=A^{(ij)}_q(t)$.

As shown in Table~\ref{tab:2}, the form factor $D_q(t)$ can be extracted from the
$(+-)$, $(-i)$ or $(ii)$ components of the tensor current using the following relations:
\begin{align}\label{eq:21}
\braket{T^{+-}_q} &= {\Gamma}_A^{+-} A_q(t) + {\Gamma}_D^{+-} D^{(+-)}_q(t),
\nonumber\\
\braket{T^{-i}_q \Delta^i} &= {\Gamma}_A^{-i} \Delta^i A_q(t) + {\Gamma}_D^{-i} \Delta^i D^{(-i)}_q(t),
\nonumber\\
\braket{T^{ii}_q} &= {\Gamma}_A^{ii} A_q(t) + {\Gamma}_D^{ii} D^{(ii)}_q(t).
\end{align}
Using  Eq.~\eqref{eq:21}, the corresponding expressions for $D_q(t)$ are given by
\begin{align}\label{eq:22}
D^{(+-)}_q(t) &= \Braket{\frac{T^{+-}_q}{{\Gamma}_D^{+-}} 
-  \frac{{\Gamma}_A^{+-}}{{\Gamma}_D^{+-}} \frac{T^{++}_q}{{\Gamma}_A^{++}}}_{\rm BT},
\nonumber\\
D^{(-i)}_q(t) &= \Braket{\frac{T^{-i}_q \Delta^i}{{\Gamma}_D^{-i} \Delta^i} 
-  \frac{{\Gamma}_A^{-i} \Delta^i}{{\Gamma}_D^{-i} \Delta^i} \frac{T^{++}_q}{{\Gamma}_A^{++}}}_{\rm BT},
\nonumber\\
D^{(ii)}_q(t) &= 
\Braket{\frac{T^{ii}_q}{{\Gamma}_D^{ii}} 
-  \frac{{\Gamma}_A^{ii}}{{\Gamma}_D^{ii}} \frac{T^{++}_q}{{\Gamma}_A^{++}}}_{\rm BT},
\end{align}
where $\frac{T^{++}_q}{{\Gamma}_A^{++}}$ is used as the input for $A_q(t)$.
Numerically, we find that $D_q(t)=D^{(+-)}_q(t)=D^{(-i)}_q(t)=D^{(ii)}_q(t)$, 
demonstrating that both $A_q(t)$ and $D_q(t)$ are extracted
in a current-component-independent manner within our self-consistent LFQM. 
This conclusion remains unchanged even when using alternative expressions for $A_q(t)$, such as 
$\frac{T^{+i}_q{\Delta^i}}{{\Gamma}_A^{+i} {\Delta^i}}$ or
$\frac{T^{ij}_q{\Delta^i}{\Delta^j}}{{\Gamma}_A^{ij} {\Delta^i}{\Delta^j}}$.

Our LFQM results satisfy four-momentum conservation, leading to the following constraints~\cite{Polyakov:2002yz}
\begin{align}\label{eq:23}
A_\pi(0) =1, \;\;  \lim_{t\to 0}t D_\pi(t)=0.
\end{align}
The second condition, known as the von Laue condition~\cite{Polyakov:2002yz}, reflects the equilibrium of internal forces within the pion. 
In addition, it has been proposed that a negative $D$-term serves as a signature of mechanical stability. 
For the pion, ChPT predicts~\cite{Donoghue:1991qv} that in the chiral limit, $D_\pi(0)=-1$. 
In general, the $D$-term receives meson mass corrections,
and a value of $D_\pi(0)=-0.97$ was predicted at $\mathcal{O}(E^2)$ in~\cite{Donoghue:1991qv}. 

The mean square radius associated with a generic form factor $F(t)$ is defined as
\begin{align}\label{eq:24}
 \Braket{r_F^2} = \frac{6}{F(0)} \frac{d F(t)}{d t} \bigg|_{t=0}.
\end{align}
Accordingly, the root-mean-square (RMS) radii of the pion's charge, mass, and mechanical ($D$-term) distributions are denoted 
as $\sqrt{\Braket{r^2_{F_\pi}}}$, $\sqrt{\Braket{r^2_{A_\pi}}}$, and $\sqrt{\Braket{r^2_{D_\pi}}}$, 
corresponding to the pion's EM form factor $F_\pi(t)$ and GFFs $A_\pi(t)$ and $D_\pi(t)$, respectively.

\section{Numerical Results}\label{Sec:IV}
In our numerical calculations, we use two sets of model parameters, $(m, \beta)=(0.22, 0.3659)$ and $(0.25, 0.3194)$ (in GeV), 
corresponding to the linear and harmonic oscillator (HO) potentials, respectively. These parameters were determined from meson mass spectrum calculations 
using the variational principle within our LFQM~\cite{CJ97}. 
While both parameter sets satisfy the constraints in Eq.~\eqref{eq:23}, neither reproduces the expected value $D_\pi(0) \approx -1 + \mathcal{O}(m_\pi^2)$.  

To address this, we identify an optimized parameter set, $(m, \beta)=(0.30, 0.28)$[GeV], which successfully satisfies this condition.
The pion decay constants obtained using the linear, HO, optimized parameter sets 
are $f_\pi=130$, $131$, and $130$ MeV, respectively. These values are in excellent agreement with the experimental value
$f_{\pi}^{\rm Exp}=131$ MeV~\cite{ParticleDataGroup:2024cfk}.

\subsection{Electromagnetic and Gravitational Form Factors}\label{Sec:IV-A}

\begin{figure}
\centering
\includegraphics[width=1.0\linewidth]{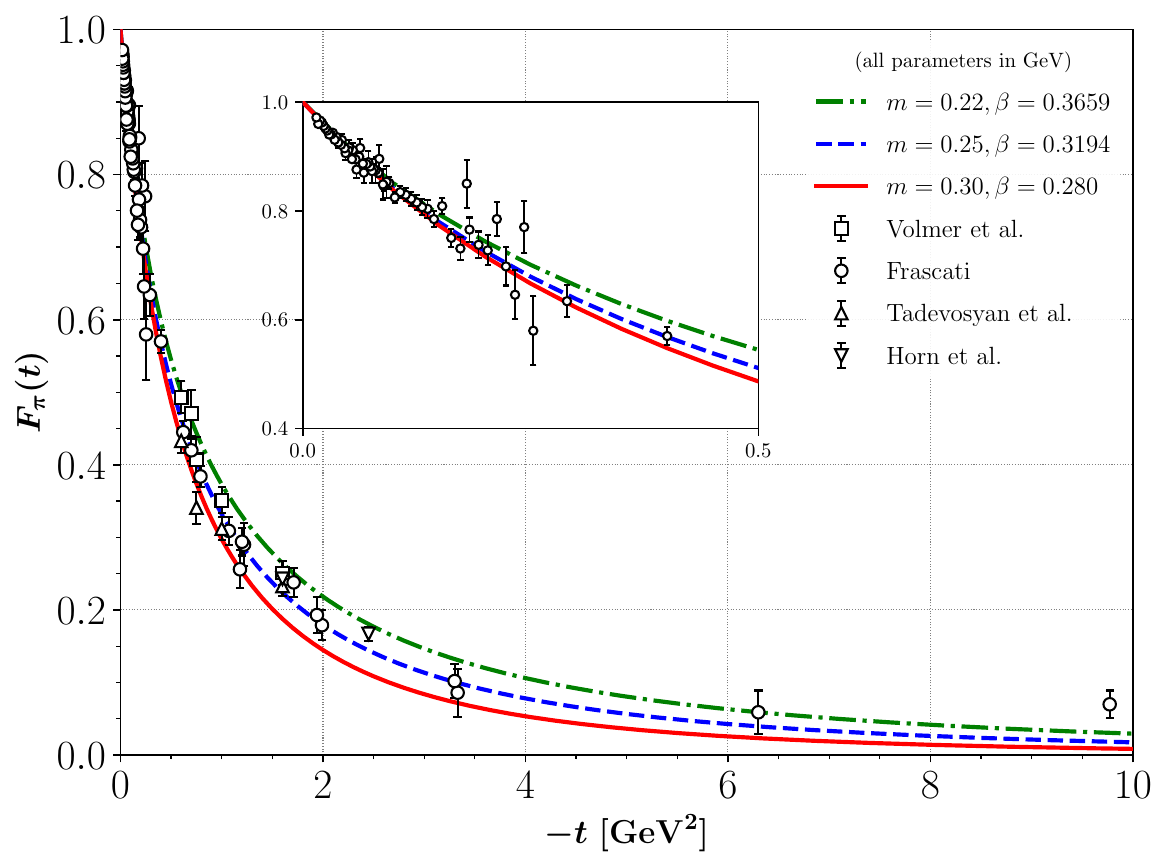}
\caption{\label{fig:1}%
Pion's EM form factor, $F_\pi(t)$, obtained from three different model parameter sets 
and compared with experimental data~\cite{Volmer2001,Amendolia,Tad2007,Horn2006}.}    
\end{figure}

Although the pion's EM form factor has been studied in previous works~\cite{CJ97,Choi:2024ptc}, we present it here for completeness,
as we employ a new parameter set, $(m, \beta)=(0.30, 0.28)$[GeV]. In~Ref.\cite{Choi:2024ptc}, it was explicitly shown that the pion EM form factor is 
independent of the current components. A more detailed analysis is provided therein.
Figure~\ref{fig:1} shows the pion EM form factor, $F_\pi(t)$, for $-t$ up to 10 ${\rm GeV}^2$, obtained using three different model parameter sets 
and compared with experimental data~\cite{Volmer2001,Amendolia,Tad2007,Horn2006}.
In the low $\vert t\vert$ region, our results exhibit good agreement with experimental data, with minimal sensitivity to the choice of parameter set. 

\begin{figure}
\centering
\includegraphics[width=1.0\linewidth]{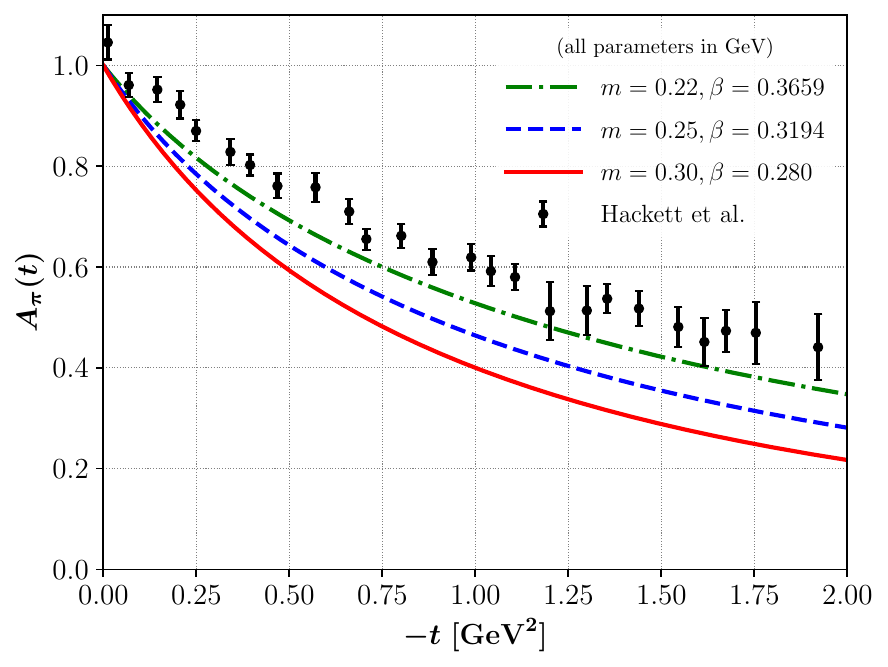}
\includegraphics[width=1.0\linewidth]{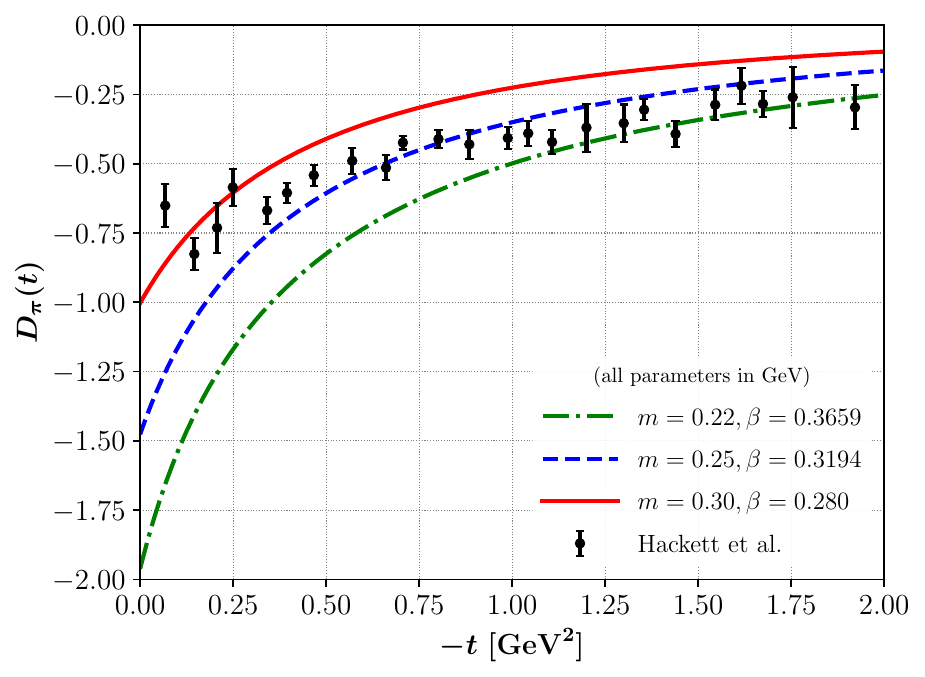}
\caption{\label{fig:2}%
Pion's GFFs, $A_\pi(t)$ (upper panel) and $D_\pi(t)$ (lower panel),
obtained from three different model parameter sets and compared with Lattice QCD results~\cite{Hackett:2023nkr}.}
\end{figure}

In Fig.~\ref{fig:2}, we present the pion's GFFs, $A_\pi(t)$ (upper panel) and $D_\pi(t)$ (lower panel), obtained using three different model parameter sets 
and compared with Lattice QCD results~\cite{Hackett:2023nkr}. As discussed in Sec.~\ref{Sec:III},
our results for the GFFs are also shown to be independent of the choice of current components.
Several observations can be drawn from Figs.~\ref{fig:1} and~\ref{fig:2}:
(1) Normalization: 
The nomalizations of $F_\pi(t)$ and $A_\pi(t)$ at $t=0$, i.e., $F_\pi(0)=A_\pi(0)=1$, 
are automatically satisfied in a parameter-independent manner.
(2) Model sensitivity of the $D$-term: 
Figure~\ref{fig:2} illustrates the sensitivity of the GFFs to the model parameters $\beta$ and $m$, 
which govern internal quark dynamics. In particular, the value of $D_\pi(0)$ shows 
the strongest dependence, with predictions of $-1.96$, $-1.48$, and $-1.00$ for the linear, HO, 
and optimized parameter sets, respectively. 
Among these, the optimized set best matches the ChPT prediction of $D_\pi(0)=-1$,
highlighting the $D$-term as a crucial constraint for refining the pion's internal structure.
The displayed curves span the full range 
of predictions allowed by parameter variations, with all sets reproducing key observables 
such as decay constant, charge radius, and electromagnetic form factors. 
The spread among the curves provides a practical estimate of the systematic uncertainties within the LFQM framework.
(3) Form factor radii ordering:
For a given parameter set, the RMS radii of the pion's charge, mass, and mechanical distributions follow the ordering
$\sqrt{\Braket{r^2_{A_\pi}}} < \sqrt{\Braket{r^2_{F_\pi}}} < \sqrt{\Braket{r^2_{D_\pi}}}$.

In Table~\ref{tab:3}, we compare the RMS radii of
the pion's charge, mass, and mechanical ($D$-term) distributions, corresponding to $F_\pi(t)$, $A_\pi(t)$, and $D_\pi(t)$, respectively,
obtained using the optimized parameter set. 
Our results for the charge and mechanical radii show good agreement with ChPT~\cite{Donoghue:1991qv}
and the experimental average reported in~\cite{ParticleDataGroup:2024cfk}. However,
the mass radius is notably larger than the values reported in earlier studies~\cite{Donoghue:1991qv,Kumano:2017lhr}. 

\begin{table}
    \caption{\label{tab:3}%
    Charge, mass, and D-term radii (in fm) of the pion from 
    theoretical, phenomenological and experimental approaches.}
        \centering
        \begin{tabular}{c@{\hskip 0.1in}c@{\hskip 0.1in}c@{\hskip 0.1in}c@{\hskip 0in}}
        \hline\hline \\ [-2ex]
        Approach & $\sqrt{\Braket{r^2_{F_\pi}}}$ & $\sqrt{\Braket{r^2_{A_\pi}}}$ & $\sqrt{\Braket{r^2_{D_\pi}}}$ 
       \\[0.5ex]
             \hline \\[-2ex]
       BT-LFWF & {0.669} & {0.546} & {0.760} \\[1.2ex]        
         ChPT($E^2$, $\mu=m_\rho$)~\cite{Donoghue:1991qv} &
        0.662 & 0.413 & 0.755 \\[1.2ex]
         MIT-LATTICE (monopole)~\cite{Hackett:2023nkr}  &
        ... & 0.41 & 0.61 \\[1.2ex]
       GDA from $\gamma^*\gamma \to \pi^0\pi^0$ \cite{Kumano:2017lhr,Belle:2015oin}&
        ... & 0.39 & 0.82 \\[1.2ex]
        EXP. Average (RPP)~\cite{ParticleDataGroup:2024cfk} &
        0.659 & ... & ... \\[1.2ex]
           \hline\hline
        \end{tabular}
    \end{table}

        

\begin{figure}
\centering
\includegraphics[width=1.0\linewidth]{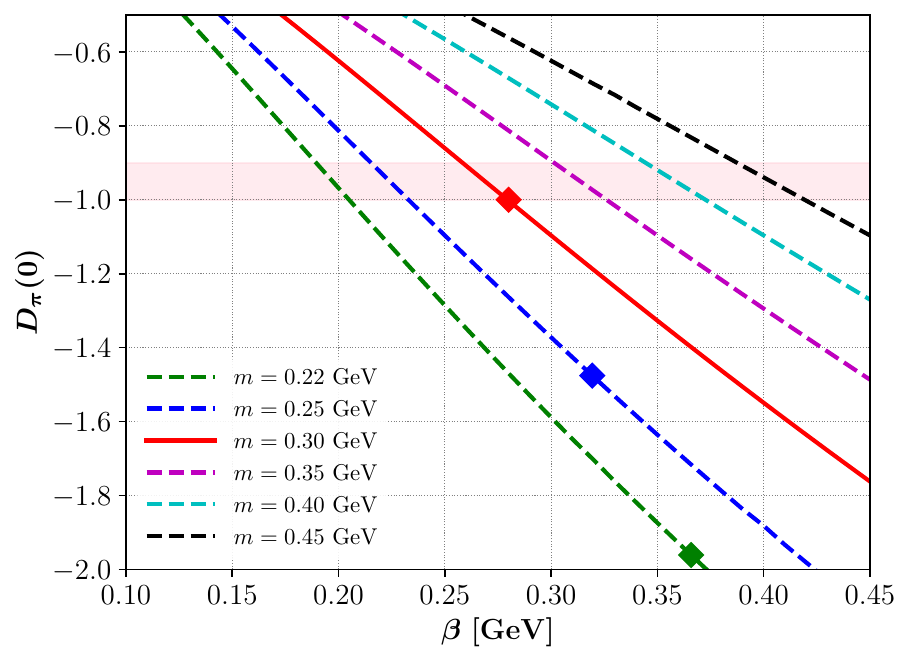}
\caption{\label{fig:3}%
Dependence of the pion's D-term, $D_\pi(0)$, on the parameters $m$ and $\beta$.}
\end{figure}

Figure~\ref{fig:3} shows the dependence of $D_\pi(0)$ on $\beta$ for different quark mass parameters. 
The red band represents the range of theoretical and lattice QCD predictions, $D_\pi(0)\in (-0.9,-1.0)$. 
The green, blue, and red squares correspond to the linear, HO, and optimized parameter sets, respectively.
Given that ChPT predicts $D_\pi(0)=-1$, we observe that in this model, the optimal value of $\beta$ increases 
monotonically with $m$ in order to satisfy this condition.

\subsection{LF zero modes from conventional LFQM}\label{Sec:IV-B}
In this subsection, we examine potential LF zero-mode contributions to the pion GFFs
that may arise when using the conventional LFQM.
As noted earlier, the conventional LFQM employs the Lorentz structure ${\Gamma}_A^{\mu\nu} = 2 P^\mu P^\nu$, 
applying the BT construction only to the matrix elements $\braket{T^{\mu\nu}_q}$, 
but not to the Lorentz structures ${\Gamma}^{\mu\nu}_{A(D)}$ themselves.
In this case, the Lorentz structures depend only on the external kinematic variables $(P^+, \bm{\Delta}_\perp, M)$ and 
can be obtained by setting $M=M'$ in Table~\ref{tab:2}, which yields $M_+=2M^2$ and $M_-=0$.

Accordingly, the form factor $A_q(t)$, extracted from the $(\mu\nu)=(++)$, $(+i)$, $(ij)$ components
in the conventional LFQM, is computed as
\begin{align}\label{eq:25}
A^{(++)}_q(t) &= \frac{\Braket{T^{++}_q}_{\rm BT}}{{\Gamma}_A^{++}}, \nonumber\\
A^{(+i)}_q(t) &= \frac{\Braket{T^{+i}_q \Delta^i}_{\rm BT}}{{\Gamma}_A^{+i} \Delta^i }, \nonumber\\
A^{(ij)}_q(t)&= \frac{\Braket{T^{ij}_q \Delta^i \Delta^j}_{\rm BT}}{{\Gamma}_A^{ij} \Delta^i \Delta^j}.
\end{align}
Similarly, the form factor $D_q(t)$, extracted from the $(\mu\nu)=(+-)$, $(-i)$, $(ii)$ components,
is given by
\begin{align}\label{eq:26}
D^{(+-)}_q(t) &=\frac{ \Braket{T^{+-}_q}_{\rm BT} }{{\Gamma}_D^{+-}} 
- \frac{{\Gamma}_A^{+-} \Braket{T^{++}_q}_{\rm BT}}{{\Gamma}_D^{+-} {\Gamma}_A^{++}},
\nonumber\\
D^{(-i)}_q(t) &= \frac{ \Braket{T^{-i}_q {\Delta}^i}_{\rm BT} }{{\Gamma}_D^{-i} {\Delta}^i_\perp} 
- \frac{{\Gamma}_A^{-i}{\Delta}^i \Braket{T^{++}_q}_{\rm BT}}{{\Gamma}_D^{-i}{\Delta}^i {\Gamma}_A^{++}}, 
\nonumber\\ 
D^{(ii)}_q(t) &= \frac{ \Braket{T^{ii}_q}_{\rm BT} }{{\Gamma}_D^{ii} } 
- \frac{{\Gamma}_A^{ii} \Braket{T^{++}_q}_{\rm BT}}{{\Gamma}_D^{ii} {\Gamma}_A^{++}}.
\end{align}

\begin{figure}
\centering
\includegraphics[width=1\linewidth]{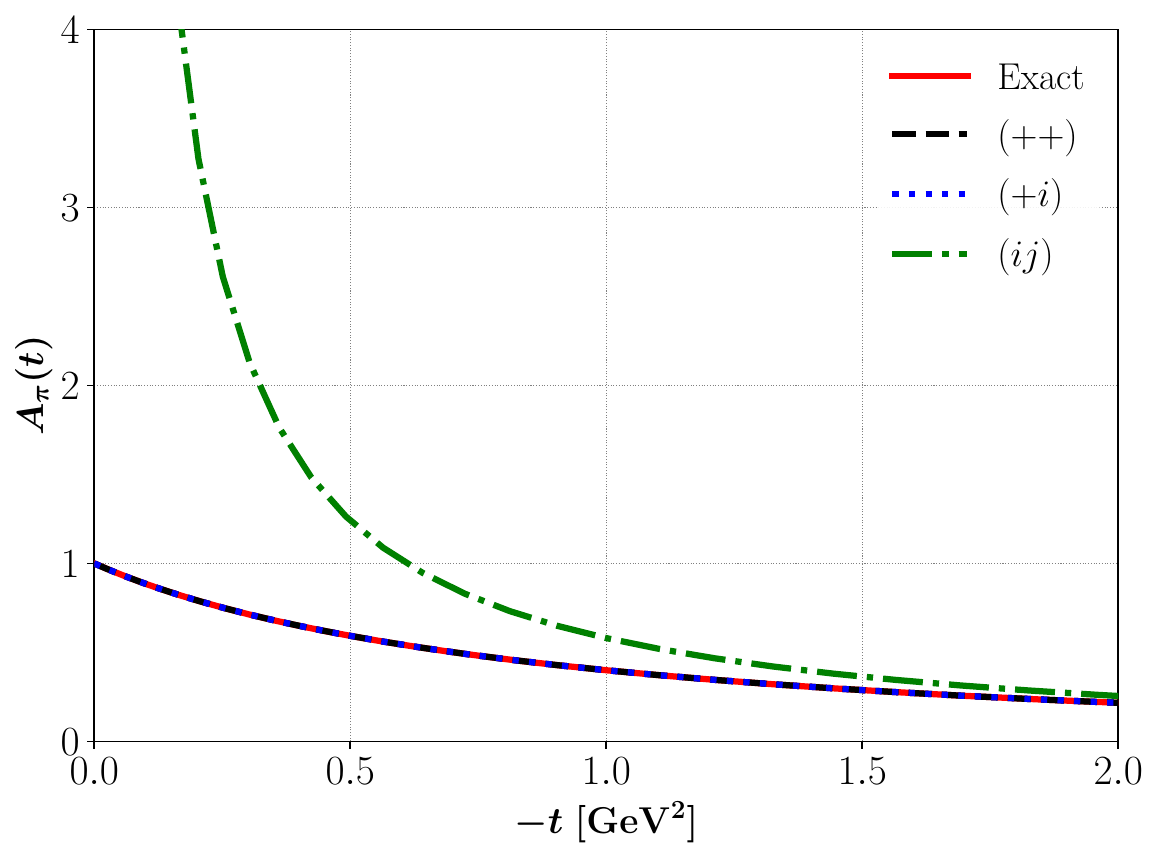}
\includegraphics[width=1\linewidth]{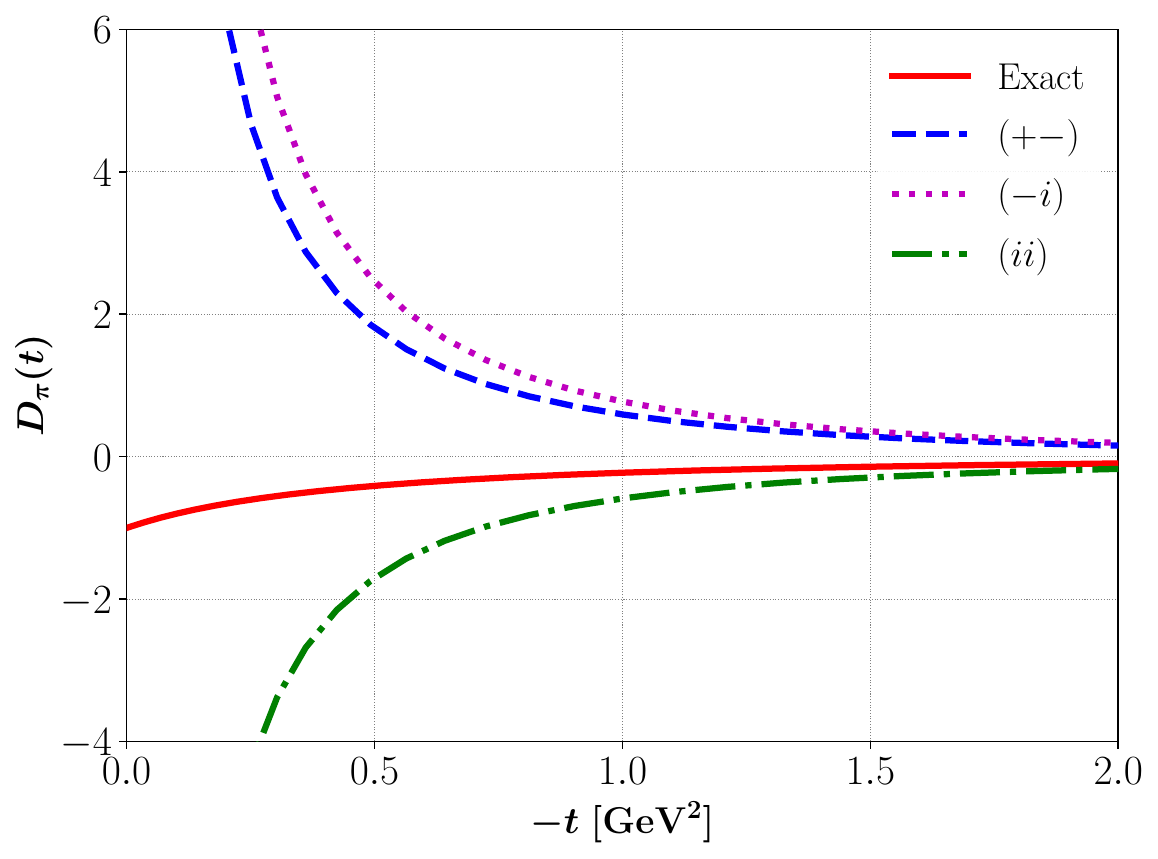}
\caption{\label{fig:4}%
Pion's GFFs $A_\pi(t)$ (top) and $D_\pi(t)$ (bottom) from conventional LFQM (with zero-mode contamination) vs. exact results obtained from BT-based LFQM.}
\end{figure}
In Fig.~\ref{fig:4}, we show the pion's GFFs, $A_\pi(t)$ (upper panel) and $D_\pi(t)$ (lower panel), obtained 
using the conventional LFQM, compared with the exact result (solid line) obtained from our BT-based LFQM. 
For consistency, the same model parameters $(m,\beta)=(0.30,0.280)$ GeV are used in both calculations. 

By comparing the results for $A_q(t)$ and $D_q(t)$ in the conventional LFQM with those obtained from
our BT-based LFQM, we find the following: The form factor $A_\pi(t)$ can also be extracted exactly in the conventional LFQM when using specific current components, 
such as $(\mu\nu)=(++)$ (dashed line) and $(+i)$ (dotted line). 
In contrast, the result from the $(ij)$ component (dot-dashed line) receives  LF zero mode contributions. The deviation from the exact result
reflects this effect, which is most pronounced near $t=0$ and gradually decreases as $-t$ increases.
For the form factor $D_\pi(t)$, all three conventional LFQM results--obtained from the $(+-)$ (dashed line), $(-i)$ (dotted line), 
and $(ii)$ (dot-dashed line) components--diverge at $t=0$, indicating that zero-mode contamination is unavoidable regardless of the current component used.
In particular, the transverse components  $(ij)$ for $A_\pi(t)$ and $(ii)$ for $D_\pi(t)$ are also shown to receive LF zero mode contributions.
These results demonstrate that exact extraction of $D_\pi(t)$ is not possible within the conventional LFQM without properly accounting for LF zero modes. 
In contrast, our BT-consistent formulation not only ensures self-consistency,
but also systematically incorporates the zero-mode contributions that conventional approaches fail to capture.

\subsection{Local densities}\label{Sec:IV-C}

The local two-dimensional (2D) LF densities associated with the GFFs provide a probabilistic interpretation of the spatial distribution of LF momentum, 
transverse pressure, and shear forces inside the pion.
These densities are based on the formalism developed in~\cite{BURKARDT_2003,Lorce:2018egm,Freese:2021czn,Freese:2021mzg},
which enables a spatially resolved analysis of the pion’s internal structure.
The LF EMT densities~\cite{Polyakov:2002yz} in the 2D transverse coordinate space are defined as~\cite{Freese:2021czn,Freese:2021mzg}
\begin{align}\label{eq:27}
T^{\mu\nu}(x_\perp) =  \int \frac{d^2 {\bm \Delta}_\perp}{(2\pi)^2}
~e^{-i {\bm \Delta}_\perp \cdot {\bm x}_\perp } 
\frac{\langle p' | \hat T^{\mu\nu}(0) | p\rangle}{\sqrt{2{p'}^{+}}\sqrt{2 p^{+}}},
\end{align}
where the 2D densities result from integrating the corresponding 3D densities over $x^{-}$,
effectively projecting them onto the transverse plane.
The relativistic normalization factor $\sqrt{2{p'}^{+}}\sqrt{2p^{+}}$ arises from the Lorentz-invariant completeness relation
and simplifies to $2P^{+}$ in the $\Delta^+=0$ frame, where $p^+=p'^+=P^+$, ensuring longitudinal momentum conservation.
Assuming localized wave functions, Eq.~\eqref{eq:27} yields physically meaningful internal EMT densities~\cite{Freese:2022fat},
allowing the matrix elements in Eq.~\eqref{eq:11} to be mapped onto spatial densities in the transverse plane.

In particular, the $T^{++}$ component corresponds to the LF momentum density $\mathcal{E}_\pi$.
The spatial components $T^{ij}$ encode the mechanical structure of the system: 
the trace part yields the isotropic pressure density $\mathcal{P}_\pi$, while the traceless part encodes
the anisotropic shear stress density $\mathcal{S}_\pi$. This decomposition allows one to isolate the 
isotropic and anisotropic components of the internal force distributions. 
Although the Fourier transform in the 2D Breit frame does not represent true rest-frame spatial densities, 
it remains a valuable tool for interpreting GFFs in terms of transverse spatial densities. 
The corresponding relations are given by~\cite{Panteleeva:2021iip,Freese:2021czn},
\begin{align}\label{eq:28}
\mathcal{E}_\pi(x_\perp) &= P^+ \tilde A_\pi(x_\perp), \nonumber\\
\mathcal{P}_\pi(x_\perp) 
&= \frac{1}{4P^+}\frac{1}{2x_\perp} \frac{d}{d x_\perp} 
\left( x_\perp  \frac{d}{d x_\perp} \tilde D_\pi(x_\perp) \right), \nonumber\\
\mathcal{S}_\pi(x_\perp) 
&= -\frac{1}{4P^+} x_\perp \frac{d}{d x_\perp} 
\left(  \frac{1}{x_\perp} \frac{d}{d x_\perp} \tilde D_\pi(x_\perp) \right).
\end{align}
where the 2D Fourier transform is defined as
\begin{align}\label{eq:29}
    \tilde {F}(x_\perp) =  \int \frac{d^2 {\bm \Delta}_\perp}{(2\pi)^2}
   \;e^{-i {\bm \Delta}_\perp \cdot {\bm x}_\perp}\; F(-{\bm \Delta}_\perp^2).
\end{align}
The LF momentum density is normalized such that its integral over the transverse plane yields the total longitudinal momentum $P^+$
of the pion~\cite{Freese:2021czn}
\begin{align}\label{eq:30}
    \int d^2x_\perp\; \mathcal{E}_\pi(x_\perp) = P^+ A_\pi(0) = P^+.
\end{align}
This condition reflects the correct momentum sum rule and ensures consistency with EMT normalization at zero momentum transfer.

The transverse mean-square radius associated with the LF momentum density can be defined through the second moment of $\mathcal{E}_\pi(x_\perp)$,
normalized by $P^+$~\cite{Freese:2021czn}:
\begin{align}\label{eq:31}
\Braket{x^2_\perp}_{\rm mom}\equiv \frac{1}{P^+}\int d^2x_\perp\; x^2_\perp \mathcal{E}_\pi(x_\perp) = 4 \frac{{\rm d}A_\pi(t)}{{\rm d}t} \bigg|_{t=0}.
\end{align}
This relation provides a spatial interpretation of the slope of the form factor $A_\pi(t)$ at zero momentum transfer.
Notably, the RMS mass radius and the RMS LF momentum radius satisfy the relation
$\Braket{r^2_{A_\pi}}^{1/2} =\sqrt{\frac{3}{2}}\Braket{x^2_\perp}_{\rm mom}^{1/2}$.
We confirm that our model satisfies the relation in Eq.~\eqref{eq:31}.

The conservation of the EMT current imposes a local stability condition that
links the pressure and shear stress densities:
\begin{align}\label{eq:32}
    \frac{d \mathcal{P}_\pi(x_\perp)}{d x_\perp} 
    + \frac{\mathcal{S}_\pi(x_\perp)}{x_\perp} + \frac{1}{2}\frac{d \mathcal{S}_\pi(x_\perp)}{d x_\perp} = 0.
\end{align}
This differential equation ensures local mechanical equilibrium within the pion.

Additionally, the 2D von Laue conditions impose global constraints:
\begin{align}\label{eq:33}
    \int d^2 x_\perp\;\mathcal{P}_\pi(x_\perp) &= 0, \nonumber\\
    \int^\infty_0 d x_\perp\;\left(\mathcal{P}_\pi(x_\perp) - \frac{1}{2}\mathcal{S}_\pi(x_\perp) \right)&= 0.
\end{align}
Equation~(\ref{eq:33}) implies that the pressure density must contain at least one node (i.e., a sign change), 
which is a necessary (though not sufficient) condition for mechanical stability.
The number of nodes depends on the spin of the hadron: 
spin-0 and spin-1/2 hadrons typically exhibit one node, while
higher-spin hadrons may exhibit more complex nodal structures.

\begin{figure}[t]
\centering
\includegraphics[width=1.0\linewidth]{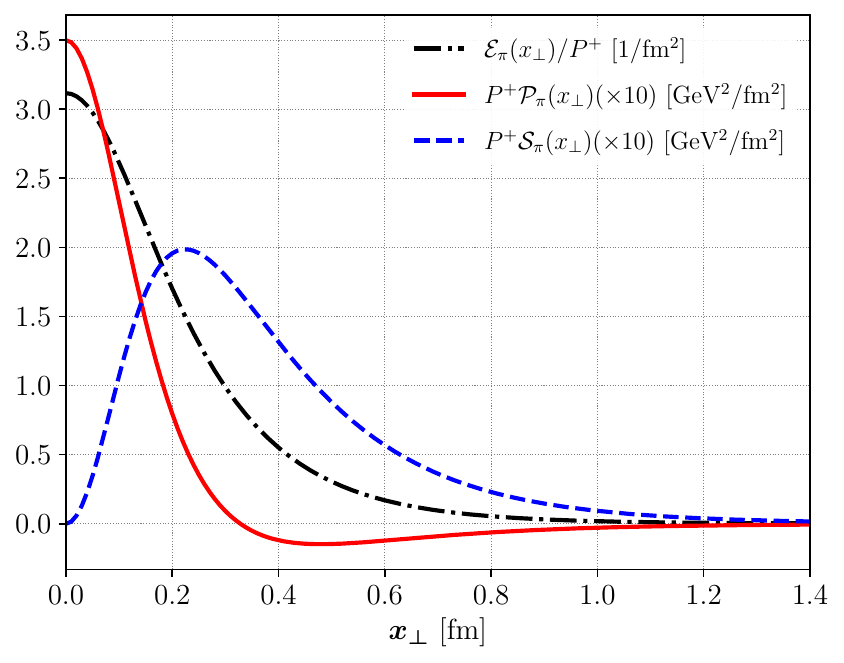}
\caption{\label{fig:5}%
2D transverse densities of the pion: $\mathcal{E}_\pi(x_\perp)$, 
$\mathcal{P}_\pi(x_\perp)$, and $\mathcal{S}_\pi(x_\perp)$. 
$\mathcal{E}_\pi$ is scaled by $1/P^+$, and $\mathcal{P}_\pi$, $\mathcal{S}_\pi$ by $10 P^+$.}    
\end{figure}

In Fig.~\ref{fig:5}, we present the 2D transverse densities of the pion obtained from $(m,\beta)=(0.30,0.280)$ GeV: 
the momentum density $\mathcal{E}_\pi(x_\perp)$ (dot-dashed line), 
the pressure density $\mathcal{P}_\pi(x_\perp)$ (solid line), and the shear stress density $\mathcal{S}_\pi(x_\perp)$ (dashed line). 
For clarity, $\mathcal{E}_\pi(x_\perp)$ is shown scaled by $1/P^+$, while $\mathcal{P}_\pi(x_\perp)$ and $\mathcal{S}_\pi(x_\perp)$ 
are scaled by $10 P^+$.

The momentum density attains its maximum at the center of the pion and monotonically decreases outward without changing sign, reflecting a uniform contribution to the LF momentum.
The pressure density is positive near the center  (up to $x_\perp = 0.33$~fm), indicating the presence of internal repulsive forces,
and becomes negative toward the periphery, signifying an attractive and stabilizing pressure.
This pressure profile closely resembles that observed in the proton~\cite{Burkert:2018bqq}.
Moreover, the pressure distribution is directly related to the $D$-term through the following expression~\cite{Freese:2021czn}
\begin{align}\label{eq:34}
    D_\pi(0) = 2P^+ \int d^2x_\perp~ x_\perp^2 \mathcal{P}_\pi(x_\perp).
\end{align}
This expression provides a quantitative measure of the internal force balance and mechanical stability of the pion. 
The shear stress density vanishes at the center ($x_\perp = 0$), reaches a maximum around $x_\perp \approx  0.2$~fm,
and remains positive throughout, without changing sign. 
Together, the pressure and shear stress densities satisfy the mechanical stability condition given in Eq.~\eqref{eq:32}.

\section{Summary}\label{Sec:V}

In this work, we have investigated the GFFs of the pion within a self-consistent LFQM 
that fully incorporates the BT construction. By  applying the BT framework uniformly to both the hadronic matrix elements and the associated Lorentz structures, 
we achieved current-component-independent extractions of the GFFs $A_\pi(t)$ and $D_\pi(t)$. 

This formulation ensures proper implementation of Lorentz covariance and plays a critical role in resolving 
LF zero-mode contributions, which otherwise obstruct consistent evaluation of 
form factors--particularly $D_\pi(t)$--in conventional LFQMs. 
In this context, inconsistency across current components often signals unaddressed zero-mode contamination. 
In our analysis, while $A_\pi(t)$ can be extracted from ``good" current components, the evaluation of $D_\pi(t)$
necessarily involves ``bad" components susceptible to zero-mode effects. 
The observed agreement across multiple current components confirms that our Lorentz-covariant BT framework 
systematically suppresses such spurious contributions, yielding physically reliable results.

By tuning the model parameters $(m, \beta)$, we identified an optimal set, $(0.30, 0.28)\,\text{GeV}$, that successfully reproduces 
the pion decay constant $f_\pi$, the EM form factor $F_\pi(t)$, and yields a $D$-term value $D_\pi(0) \approx -1$, 
in agreement with predictions from ChPT.
The $D$-term plays a pivotal role in characterizing the pion's internal structure, particularly its mechanical properties. 
Unlike $F_\pi(t)$ and $A_\pi(t)$, which are well-constrained and parameter-independent at $t=0$,
the value of $D_\pi(0)$ is highly sensitive to the model parameters. 
This sensitivity makes it a powerful probe for refining the pion’s wave function and internal dynamics. 
Moreover, the $D$-term governs the mechanical radius of the pion, which is found to be the largest among the charge, mass, and mechanical radii, 
highlighting its significance in understanding the spatial distribution of internal forces within the hadron.

We also conducted a detailed analysis of the pion’s transverse structure through the 2D  
LF densities associated with the GFFs: the LF momentum density $\mathcal{E}_\pi(x_\perp)$, 
transverse pressure density $\mathcal{P}_\pi(x_\perp)$, and shear stress density $\mathcal{S}_\pi(x_\perp)$. 
These densities satisfy the required normalization conditions 
and mechanical stability relations, including the von Laue constraints. 

Our results show that the momentum density peaks at the center of the pion and decreases monotonically
without changing sign, reflecting its uniform contribution to the LF momentum.
The pressure density is positive near the center (up to $x_\perp = 0.33$~fm), indicating internal repulsion, and becomes negative toward the periphery, 
signifying an attractive and stabilizing force. The shear stress peaks at an intermediate distance around $x_\perp \approx  0.2$~fm
and remains positive throughout, complementing the pressure profile to maintain mechanical equilibrium. 

Overall, our BT-based LFQM provides a robust, zero-mode free, and current-component-independent framework for analyzing 
both the EM form factors and GFFs of the pion.
Future work may extend this approach to other hadrons and explore the scale evolution of GFFs and associated densities within a QCD-consistent framework.


\section*{Acknowledgements}
The present work was supported by the National Research Foundation of Korea(NRF) grant funded by the Korea government(MSIT) under Grant No. RS-2025-00513982(Y. Choi), RS-2023-00210298(H.-D. Son), RS-2023-NR076506(H.-M. Choi).

\bibliographystyle{apsrev4-2}
\bibliography{references}

\begin{thebibliography}{69}%
\makeatletter
\providecommand \@ifxundefined [1]{%
 \@ifx{#1\undefined}
}%
\providecommand \@ifnum [1]{%
 \ifnum #1\expandafter \@firstoftwo
 \else \expandafter \@secondoftwo
 \fi
}%
\providecommand \@ifx [1]{%
 \ifx #1\expandafter \@firstoftwo
 \else \expandafter \@secondoftwo
 \fi
}%
\providecommand \natexlab [1]{#1}%
\providecommand \enquote  [1]{``#1''}%
\providecommand \bibnamefont  [1]{#1}%
\providecommand \bibfnamefont [1]{#1}%
\providecommand \citenamefont [1]{#1}%
\providecommand \href@noop [0]{\@secondoftwo}%
\providecommand \href [0]{\begingroup \@sanitize@url \@href}%
\providecommand \@href[1]{\@@startlink{#1}\@@href}%
\providecommand \@@href[1]{\endgroup#1\@@endlink}%
\providecommand \@sanitize@url [0]{\catcode `\\12\catcode `\$12\catcode `\&12\catcode `\#12\catcode `\^12\catcode `\_12\catcode `\%12\relax}%
\providecommand \@@startlink[1]{}%
\providecommand \@@endlink[0]{}%
\providecommand \url  [0]{\begingroup\@sanitize@url \@url }%
\providecommand \@url [1]{\endgroup\@href {#1}{\urlprefix }}%
\providecommand \urlprefix  [0]{URL }%
\providecommand \Eprint [0]{\href }%
\providecommand \doibase [0]{https://doi.org/}%
\providecommand \selectlanguage [0]{\@gobble}%
\providecommand \bibinfo  [0]{\@secondoftwo}%
\providecommand \bibfield  [0]{\@secondoftwo}%
\providecommand \translation [1]{[#1]}%
\providecommand \BibitemOpen [0]{}%
\providecommand \bibitemStop [0]{}%
\providecommand \bibitemNoStop [0]{.\EOS\space}%
\providecommand \EOS [0]{\spacefactor3000\relax}%
\providecommand \BibitemShut  [1]{\csname bibitem#1\endcsname}%
\let\auto@bib@innerbib\@empty
\bibitem [{\citenamefont {Kobzarev}\ and\ \citenamefont {Okun}(1962)}]{Kobzarev:1962wt}%
  \BibitemOpen
  \bibfield  {author} {\bibinfo {author} {\bibfnamefont {I.~Y.}\ \bibnamefont {Kobzarev}}\ and\ \bibinfo {author} {\bibfnamefont {L.~B.}\ \bibnamefont {Okun}},\ }\href@noop {} {\bibfield  {journal} {\bibinfo  {journal} {Zh. Eksp. Teor. Fiz.}\ }\textbf {\bibinfo {volume} {43}},\ \bibinfo {pages} {1904} (\bibinfo {year} {1962})}\BibitemShut {NoStop}%
\bibitem [{\citenamefont {Pagels}(1966)}]{Pagels:1966zza}%
  \BibitemOpen
  \bibfield  {author} {\bibinfo {author} {\bibfnamefont {H.}~\bibnamefont {Pagels}},\ }\href {https://doi.org/10.1103/PhysRev.144.1250} {\bibfield  {journal} {\bibinfo  {journal} {Phys. Rev.}\ }\textbf {\bibinfo {volume} {144}},\ \bibinfo {pages} {1250} (\bibinfo {year} {1966})}\BibitemShut {NoStop}%
\bibitem [{\citenamefont {Polyakov}(2003)}]{Polyakov:2002yz}%
  \BibitemOpen
  \bibfield  {author} {\bibinfo {author} {\bibfnamefont {M.~V.}\ \bibnamefont {Polyakov}},\ }\href {https://doi.org/10.1016/S0370-2693(03)00036-4} {\bibfield  {journal} {\bibinfo  {journal} {Phys. Lett. B}\ }\textbf {\bibinfo {volume} {555}},\ \bibinfo {pages} {57} (\bibinfo {year} {2003})},\ \Eprint {https://arxiv.org/abs/hep-ph/0210165} {arXiv:hep-ph/0210165} \BibitemShut {NoStop}%
\bibitem [{\citenamefont {Lorc\'e}\ \emph {et~al.}(2019)\citenamefont {Lorc\'e}, \citenamefont {Moutarde},\ and\ \citenamefont {Trawi\'nski}}]{Lorce:2018egm}%
  \BibitemOpen
  \bibfield  {author} {\bibinfo {author} {\bibfnamefont {C.}~\bibnamefont {Lorc\'e}}, \bibinfo {author} {\bibfnamefont {H.}~\bibnamefont {Moutarde}},\ and\ \bibinfo {author} {\bibfnamefont {A.~P.}\ \bibnamefont {Trawi\'nski}},\ }\href {https://doi.org/10.1140/epjc/s10052-019-6572-3} {\bibfield  {journal} {\bibinfo  {journal} {Eur. Phys. J. C}\ }\textbf {\bibinfo {volume} {79}},\ \bibinfo {pages} {89} (\bibinfo {year} {2019})},\ \Eprint {https://arxiv.org/abs/1810.09837} {arXiv:1810.09837 [hep-ph]} \BibitemShut {NoStop}%
\bibitem [{\citenamefont {Panteleeva}\ and\ \citenamefont {Polyakov}(2020)}]{Panteleeva:2020ejw}%
  \BibitemOpen
  \bibfield  {author} {\bibinfo {author} {\bibfnamefont {J.~Y.}\ \bibnamefont {Panteleeva}}\ and\ \bibinfo {author} {\bibfnamefont {M.~V.}\ \bibnamefont {Polyakov}},\ }\href {https://doi.org/10.1016/j.physletb.2020.135707} {\bibfield  {journal} {\bibinfo  {journal} {Phys. Lett. B}\ }\textbf {\bibinfo {volume} {809}},\ \bibinfo {pages} {135707} (\bibinfo {year} {2020})},\ \Eprint {https://arxiv.org/abs/2004.02912} {arXiv:2004.02912 [hep-ph]} \BibitemShut {NoStop}%
\bibitem [{\citenamefont {Kim}\ \emph {et~al.}(2021)\citenamefont {Kim}, \citenamefont {Kim}, \citenamefont {Polyakov},\ and\ \citenamefont {Son}}]{Kim:2020nug}%
  \BibitemOpen
  \bibfield  {author} {\bibinfo {author} {\bibfnamefont {J.-Y.}\ \bibnamefont {Kim}}, \bibinfo {author} {\bibfnamefont {H.-C.}\ \bibnamefont {Kim}}, \bibinfo {author} {\bibfnamefont {M.~V.}\ \bibnamefont {Polyakov}},\ and\ \bibinfo {author} {\bibfnamefont {H.-D.}\ \bibnamefont {Son}},\ }\href {https://doi.org/10.1103/PhysRevD.103.014015} {\bibfield  {journal} {\bibinfo  {journal} {Phys. Rev. D}\ }\textbf {\bibinfo {volume} {103}},\ \bibinfo {pages} {014015} (\bibinfo {year} {2021})},\ \Eprint {https://arxiv.org/abs/2008.06652} {arXiv:2008.06652 [hep-ph]} \BibitemShut {NoStop}%
\bibitem [{\citenamefont {Polyakov}\ and\ \citenamefont {Schweitzer}(2018)}]{Polyakov:2018zvc}%
  \BibitemOpen
  \bibfield  {author} {\bibinfo {author} {\bibfnamefont {M.~V.}\ \bibnamefont {Polyakov}}\ and\ \bibinfo {author} {\bibfnamefont {P.}~\bibnamefont {Schweitzer}},\ }\href {https://doi.org/10.1142/S0217751X18300259} {\bibfield  {journal} {\bibinfo  {journal} {Int. J. Mod. Phys. A}\ }\textbf {\bibinfo {volume} {33}},\ \bibinfo {pages} {1830025} (\bibinfo {year} {2018})},\ \Eprint {https://arxiv.org/abs/1805.06596} {arXiv:1805.06596 [hep-ph]} \BibitemShut {NoStop}%
\bibitem [{\citenamefont {Lorc\'e}\ and\ \citenamefont {Schweitzer}(2025)}]{Lorce:2025oot}%
  \BibitemOpen
  \bibfield  {author} {\bibinfo {author} {\bibfnamefont {C.}~\bibnamefont {Lorc\'e}}\ and\ \bibinfo {author} {\bibfnamefont {P.}~\bibnamefont {Schweitzer}},\ }\href@noop {} {\bibfield  {journal} {\bibinfo  {journal} {~}\ } (\bibinfo {year} {2025})},\ \Eprint {https://arxiv.org/abs/2501.04622} {arXiv:2501.04622 [hep-ph]} \BibitemShut {NoStop}%
\bibitem [{\citenamefont {M\"uller}\ \emph {et~al.}(1994)\citenamefont {M\"uller}, \citenamefont {Robaschik}, \citenamefont {Geyer}, \citenamefont {Dittes},\ and\ \citenamefont {Ho\v{r}ej\v{s}i}}]{Muller:1994ses}%
  \BibitemOpen
  \bibfield  {author} {\bibinfo {author} {\bibfnamefont {D.}~\bibnamefont {M\"uller}}, \bibinfo {author} {\bibfnamefont {D.}~\bibnamefont {Robaschik}}, \bibinfo {author} {\bibfnamefont {B.}~\bibnamefont {Geyer}}, \bibinfo {author} {\bibfnamefont {F.~M.}\ \bibnamefont {Dittes}},\ and\ \bibinfo {author} {\bibfnamefont {J.}~\bibnamefont {Ho\v{r}ej\v{s}i}},\ }\href {https://doi.org/10.1002/prop.2190420202} {\bibfield  {journal} {\bibinfo  {journal} {Fortsch. Phys.}\ }\textbf {\bibinfo {volume} {42}},\ \bibinfo {pages} {101} (\bibinfo {year} {1994})},\ \Eprint {https://arxiv.org/abs/hep-ph/9812448} {arXiv:hep-ph/9812448} \BibitemShut {NoStop}%
\bibitem [{\citenamefont {Burkert}\ \emph {et~al.}(2018)\citenamefont {Burkert}, \citenamefont {Elouadrhiri},\ and\ \citenamefont {Girod}}]{Burkert:2018bqq}%
  \BibitemOpen
  \bibfield  {author} {\bibinfo {author} {\bibfnamefont {V.~D.}\ \bibnamefont {Burkert}}, \bibinfo {author} {\bibfnamefont {L.}~\bibnamefont {Elouadrhiri}},\ and\ \bibinfo {author} {\bibfnamefont {F.~X.}\ \bibnamefont {Girod}},\ }\href {https://doi.org/10.1038/s41586-018-0060-z} {\bibfield  {journal} {\bibinfo  {journal} {Nature}\ }\textbf {\bibinfo {volume} {557}},\ \bibinfo {pages} {396} (\bibinfo {year} {2018})}\BibitemShut {NoStop}%
\bibitem [{\citenamefont {Duran}\ \emph {et~al.}(2023)\citenamefont {Duran} \emph {et~al.}}]{Duran:2022xag}%
  \BibitemOpen
  \bibfield  {author} {\bibinfo {author} {\bibfnamefont {B.}~\bibnamefont {Duran}} \emph {et~al.},\ }\href {https://doi.org/10.1038/s41586-023-05730-4} {\bibfield  {journal} {\bibinfo  {journal} {Nature}\ }\textbf {\bibinfo {volume} {615}},\ \bibinfo {pages} {813} (\bibinfo {year} {2023})},\ \Eprint {https://arxiv.org/abs/2207.05212} {arXiv:2207.05212 [nucl-ex]} \BibitemShut {NoStop}%
\bibitem [{\citenamefont {Adhikari}\ \emph {et~al.}(2023)\citenamefont {Adhikari} \emph {et~al.}}]{GlueX:2023pev}%
  \BibitemOpen
  \bibfield  {author} {\bibinfo {author} {\bibfnamefont {S.}~\bibnamefont {Adhikari}} \emph {et~al.} (\bibinfo {collaboration} {GlueX}),\ }\href {https://doi.org/10.1103/PhysRevC.108.025201} {\bibfield  {journal} {\bibinfo  {journal} {Phys. Rev. C}\ }\textbf {\bibinfo {volume} {108}},\ \bibinfo {pages} {025201} (\bibinfo {year} {2023})},\ \Eprint {https://arxiv.org/abs/2304.03845} {arXiv:2304.03845 [nucl-ex]} \BibitemShut {NoStop}%
\bibitem [{\citenamefont {Guo}\ \emph {et~al.}(2023)\citenamefont {Guo}, \citenamefont {Ji}, \citenamefont {Liu},\ and\ \citenamefont {Yang}}]{Guo:2023pqw}%
  \BibitemOpen
  \bibfield  {author} {\bibinfo {author} {\bibfnamefont {Y.}~\bibnamefont {Guo}}, \bibinfo {author} {\bibfnamefont {X.}~\bibnamefont {Ji}}, \bibinfo {author} {\bibfnamefont {Y.}~\bibnamefont {Liu}},\ and\ \bibinfo {author} {\bibfnamefont {J.}~\bibnamefont {Yang}},\ }\href {https://doi.org/10.1103/PhysRevD.108.034003} {\bibfield  {journal} {\bibinfo  {journal} {Phys. Rev. D}\ }\textbf {\bibinfo {volume} {108}},\ \bibinfo {pages} {034003} (\bibinfo {year} {2023})},\ \Eprint {https://arxiv.org/abs/2305.06992} {arXiv:2305.06992 [hep-ph]} \BibitemShut {NoStop}%
\bibitem [{\citenamefont {Son}\ and\ \citenamefont {Kim}(2014)}]{Son:2014sna}%
  \BibitemOpen
  \bibfield  {author} {\bibinfo {author} {\bibfnamefont {H.-D.}\ \bibnamefont {Son}}\ and\ \bibinfo {author} {\bibfnamefont {H.-C.}\ \bibnamefont {Kim}},\ }\href {https://doi.org/10.1103/PhysRevD.90.111901} {\bibfield  {journal} {\bibinfo  {journal} {Phys. Rev. D}\ }\textbf {\bibinfo {volume} {90}},\ \bibinfo {pages} {111901} (\bibinfo {year} {2014})},\ \Eprint {https://arxiv.org/abs/1410.1420} {arXiv:1410.1420 [hep-ph]} \BibitemShut {NoStop}%
\bibitem [{\citenamefont {Kumano}\ \emph {et~al.}(2018)\citenamefont {Kumano}, \citenamefont {Song},\ and\ \citenamefont {Teryaev}}]{Kumano:2017lhr}%
  \BibitemOpen
  \bibfield  {author} {\bibinfo {author} {\bibfnamefont {S.}~\bibnamefont {Kumano}}, \bibinfo {author} {\bibfnamefont {Q.-T.}\ \bibnamefont {Song}},\ and\ \bibinfo {author} {\bibfnamefont {O.~V.}\ \bibnamefont {Teryaev}},\ }\href {https://doi.org/10.1103/PhysRevD.97.014020} {\bibfield  {journal} {\bibinfo  {journal} {Phys. Rev. D}\ }\textbf {\bibinfo {volume} {97}},\ \bibinfo {pages} {014020} (\bibinfo {year} {2018})},\ \Eprint {https://arxiv.org/abs/1711.08088} {arXiv:1711.08088 [hep-ph]} \BibitemShut {NoStop}%
\bibitem [{\citenamefont {Masuda}\ \emph {et~al.}(2016)\citenamefont {Masuda} \emph {et~al.}}]{Belle:2015oin}%
  \BibitemOpen
  \bibfield  {author} {\bibinfo {author} {\bibfnamefont {M.}~\bibnamefont {Masuda}} \emph {et~al.} (\bibinfo {collaboration} {Belle}),\ }\href {https://doi.org/10.1103/PhysRevD.93.032003} {\bibfield  {journal} {\bibinfo  {journal} {Phys. Rev. D}\ }\textbf {\bibinfo {volume} {93}},\ \bibinfo {pages} {032003} (\bibinfo {year} {2016})},\ \Eprint {https://arxiv.org/abs/1508.06757} {arXiv:1508.06757 [hep-ex]} \BibitemShut {NoStop}%
\bibitem [{\citenamefont {Amrath}\ \emph {et~al.}(2008)\citenamefont {Amrath}, \citenamefont {Diehl},\ and\ \citenamefont {Lansberg}}]{Amrath:2008vx}%
  \BibitemOpen
  \bibfield  {author} {\bibinfo {author} {\bibfnamefont {D.}~\bibnamefont {Amrath}}, \bibinfo {author} {\bibfnamefont {M.}~\bibnamefont {Diehl}},\ and\ \bibinfo {author} {\bibfnamefont {J.-P.}\ \bibnamefont {Lansberg}},\ }\href {https://doi.org/10.1140/epjc/s10052-008-0769-1} {\bibfield  {journal} {\bibinfo  {journal} {Eur. Phys. J. C}\ }\textbf {\bibinfo {volume} {58}},\ \bibinfo {pages} {179} (\bibinfo {year} {2008})},\ \Eprint {https://arxiv.org/abs/0807.4474} {arXiv:0807.4474 [hep-ph]} \BibitemShut {NoStop}%
\bibitem [{\citenamefont {Ch\'avez}\ \emph {et~al.}(2022)\citenamefont {Ch\'avez}, \citenamefont {Bertone}, \citenamefont {De~Soto~Borrero}, \citenamefont {Defurne}, \citenamefont {Mezrag}, \citenamefont {Moutarde}, \citenamefont {Rodr\'\i{}guez-Quintero},\ and\ \citenamefont {Segovia}}]{Chavez:2021koz}%
  \BibitemOpen
  \bibfield  {author} {\bibinfo {author} {\bibfnamefont {J.~M.~M.}\ \bibnamefont {Ch\'avez}}, \bibinfo {author} {\bibfnamefont {V.}~\bibnamefont {Bertone}}, \bibinfo {author} {\bibfnamefont {F.}~\bibnamefont {De~Soto~Borrero}}, \bibinfo {author} {\bibfnamefont {M.}~\bibnamefont {Defurne}}, \bibinfo {author} {\bibfnamefont {C.}~\bibnamefont {Mezrag}}, \bibinfo {author} {\bibfnamefont {H.}~\bibnamefont {Moutarde}}, \bibinfo {author} {\bibfnamefont {J.}~\bibnamefont {Rodr\'\i{}guez-Quintero}},\ and\ \bibinfo {author} {\bibfnamefont {J.}~\bibnamefont {Segovia}},\ }\href {https://doi.org/10.1103/PhysRevLett.128.202501} {\bibfield  {journal} {\bibinfo  {journal} {Phys. Rev. Lett.}\ }\textbf {\bibinfo {volume} {128}},\ \bibinfo {pages} {202501} (\bibinfo {year} {2022})},\ \Eprint {https://arxiv.org/abs/2110.09462} {arXiv:2110.09462 [hep-ph]} \BibitemShut {NoStop}%
\bibitem [{\citenamefont {Hatta}\ and\ \citenamefont {Schoenleber}(2025)}]{Hatta:2025ryj}%
  \BibitemOpen
  \bibfield  {author} {\bibinfo {author} {\bibfnamefont {Y.}~\bibnamefont {Hatta}}\ and\ \bibinfo {author} {\bibfnamefont {J.}~\bibnamefont {Schoenleber}},\ }\href@noop {} {\bibfield  {journal} {\bibinfo  {journal} {~}\ } (\bibinfo {year} {2025})},\ \Eprint {https://arxiv.org/abs/2502.12061} {arXiv:2502.12061 [hep-ph]} \BibitemShut {NoStop}%
\bibitem [{\citenamefont {Brommel}(2007)}]{Brommel:2007zz}%
  \BibitemOpen
  \bibfield  {author} {\bibinfo {author} {\bibfnamefont {D.}~\bibnamefont {Brommel}},\ }\emph {\bibinfo {title} {{Pion Structure from the Lattice}}},\ \href {https://doi.org/10.3204/DESY-THESIS-2007-023} {Ph.D. thesis},\ \bibinfo  {school} {Regensburg U.} (\bibinfo {year} {2007})\BibitemShut {NoStop}%
\bibitem [{\citenamefont {Br\"ommel}\ \emph {et~al.}(2008)\citenamefont {Br\"ommel} \emph {et~al.}}]{QCDSF:2007ifr}%
  \BibitemOpen
  \bibfield  {author} {\bibinfo {author} {\bibfnamefont {D.}~\bibnamefont {Br\"ommel}} \emph {et~al.} (\bibinfo {collaboration} {QCDSF, UKQCD}),\ }\href {https://doi.org/10.1103/PhysRevLett.101.122001} {\bibfield  {journal} {\bibinfo  {journal} {Phys. Rev. Lett.}\ }\textbf {\bibinfo {volume} {101}},\ \bibinfo {pages} {122001} (\bibinfo {year} {2008})},\ \Eprint {https://arxiv.org/abs/0708.2249} {arXiv:0708.2249 [hep-lat]} \BibitemShut {NoStop}%
\bibitem [{\citenamefont {Shanahan}\ and\ \citenamefont {Detmold}(2019)}]{Shanahan:2018pib}%
  \BibitemOpen
  \bibfield  {author} {\bibinfo {author} {\bibfnamefont {P.~E.}\ \bibnamefont {Shanahan}}\ and\ \bibinfo {author} {\bibfnamefont {W.}~\bibnamefont {Detmold}},\ }\href {https://doi.org/10.1103/PhysRevD.99.014511} {\bibfield  {journal} {\bibinfo  {journal} {Phys. Rev. D}\ }\textbf {\bibinfo {volume} {99}},\ \bibinfo {pages} {014511} (\bibinfo {year} {2019})},\ \Eprint {https://arxiv.org/abs/1810.04626} {arXiv:1810.04626 [hep-lat]} \BibitemShut {NoStop}%
\bibitem [{\citenamefont {Pefkou}\ \emph {et~al.}(2022)\citenamefont {Pefkou}, \citenamefont {Hackett},\ and\ \citenamefont {Shanahan}}]{Pefkou:2021fni}%
  \BibitemOpen
  \bibfield  {author} {\bibinfo {author} {\bibfnamefont {D.~A.}\ \bibnamefont {Pefkou}}, \bibinfo {author} {\bibfnamefont {D.~C.}\ \bibnamefont {Hackett}},\ and\ \bibinfo {author} {\bibfnamefont {P.~E.}\ \bibnamefont {Shanahan}},\ }\href {https://doi.org/10.1103/PhysRevD.105.054509} {\bibfield  {journal} {\bibinfo  {journal} {Phys. Rev. D}\ }\textbf {\bibinfo {volume} {105}},\ \bibinfo {pages} {054509} (\bibinfo {year} {2022})},\ \Eprint {https://arxiv.org/abs/2107.10368} {arXiv:2107.10368 [hep-lat]} \BibitemShut {NoStop}%
\bibitem [{\citenamefont {Hackett}\ \emph {et~al.}(2023)\citenamefont {Hackett}, \citenamefont {Oare}, \citenamefont {Pefkou},\ and\ \citenamefont {Shanahan}}]{Hackett:2023nkr}%
  \BibitemOpen
  \bibfield  {author} {\bibinfo {author} {\bibfnamefont {D.~C.}\ \bibnamefont {Hackett}}, \bibinfo {author} {\bibfnamefont {P.~R.}\ \bibnamefont {Oare}}, \bibinfo {author} {\bibfnamefont {D.~A.}\ \bibnamefont {Pefkou}},\ and\ \bibinfo {author} {\bibfnamefont {P.~E.}\ \bibnamefont {Shanahan}},\ }\href {https://doi.org/10.1103/PhysRevD.108.114504} {\bibfield  {journal} {\bibinfo  {journal} {Phys. Rev. D}\ }\textbf {\bibinfo {volume} {108}},\ \bibinfo {pages} {114504} (\bibinfo {year} {2023})},\ \Eprint {https://arxiv.org/abs/2307.11707} {arXiv:2307.11707 [hep-lat]} \BibitemShut {NoStop}%
\bibitem [{\citenamefont {Donoghue}\ and\ \citenamefont {Leutwyler}(1991)}]{Donoghue:1991qv}%
  \BibitemOpen
  \bibfield  {author} {\bibinfo {author} {\bibfnamefont {J.~F.}\ \bibnamefont {Donoghue}}\ and\ \bibinfo {author} {\bibfnamefont {H.}~\bibnamefont {Leutwyler}},\ }\href {https://doi.org/10.1007/BF01560453} {\bibfield  {journal} {\bibinfo  {journal} {Z. Phys. C}\ }\textbf {\bibinfo {volume} {52}},\ \bibinfo {pages} {343} (\bibinfo {year} {1991})}\BibitemShut {NoStop}%
\bibitem [{\citenamefont {Broniowski}\ and\ \citenamefont {Arriola}(2008)}]{WB08}%
  \BibitemOpen
  \bibfield  {author} {\bibinfo {author} {\bibfnamefont {W.}~\bibnamefont {Broniowski}}\ and\ \bibinfo {author} {\bibfnamefont {E.~R.}\ \bibnamefont {Arriola}},\ }\href {https://doi.org/10.1103/PhysRevD.78.094011} {\bibfield  {journal} {\bibinfo  {journal} {Phys. Rev. D}\ }\textbf {\bibinfo {volume} {78}},\ \bibinfo {pages} {094011} (\bibinfo {year} {2008})}\BibitemShut {NoStop}%
\bibitem [{\citenamefont {Freese}\ and\ \citenamefont {Clo\"et}(2019)}]{Freese_2019}%
  \BibitemOpen
  \bibfield  {author} {\bibinfo {author} {\bibfnamefont {A.}~\bibnamefont {Freese}}\ and\ \bibinfo {author} {\bibfnamefont {I.~C.}\ \bibnamefont {Clo\"et}},\ }\href {https://doi.org/10.1103/PhysRevC.100.015201} {\bibfield  {journal} {\bibinfo  {journal} {Phys. Rev. C}\ }\textbf {\bibinfo {volume} {100}},\ \bibinfo {pages} {015201} (\bibinfo {year} {2019})}\BibitemShut {NoStop}%
\bibitem [{\citenamefont {Krutov}\ and\ \citenamefont {Troitsky}(2021)}]{Krutov21}%
  \BibitemOpen
  \bibfield  {author} {\bibinfo {author} {\bibfnamefont {A.~F.}\ \bibnamefont {Krutov}}\ and\ \bibinfo {author} {\bibfnamefont {V.~E.}\ \bibnamefont {Troitsky}},\ }\href {https://doi.org/10.1103/PhysRevD.103.014029} {\bibfield  {journal} {\bibinfo  {journal} {Phys. Rev. D}\ }\textbf {\bibinfo {volume} {103}},\ \bibinfo {pages} {014029} (\bibinfo {year} {2021})}\BibitemShut {NoStop}%
\bibitem [{\citenamefont {Xing}\ \emph {et~al.}(2023)\citenamefont {Xing}, \citenamefont {Ding},\ and\ \citenamefont {Chang}}]{ZXing}%
  \BibitemOpen
  \bibfield  {author} {\bibinfo {author} {\bibfnamefont {Z.}~\bibnamefont {Xing}}, \bibinfo {author} {\bibfnamefont {M.}~\bibnamefont {Ding}},\ and\ \bibinfo {author} {\bibfnamefont {L.}~\bibnamefont {Chang}},\ }\href {https://doi.org/10.1103/PhysRevD.107.L031502} {\bibfield  {journal} {\bibinfo  {journal} {Phys. Rev. D}\ }\textbf {\bibinfo {volume} {107}},\ \bibinfo {pages} {L031502} (\bibinfo {year} {2023})}\BibitemShut {NoStop}%
\bibitem [{\citenamefont {{Xu}}\ \emph {et~al.}(2024)\citenamefont {{Xu}}, \citenamefont {{Ding}}, \citenamefont {{Raya}}, \citenamefont {{Roberts}}, \citenamefont {{Rodr{\'\i}guez-Quintero}},\ and\ \citenamefont {{Schmidt}}}]{xu2023}%
  \BibitemOpen
  \bibfield  {author} {\bibinfo {author} {\bibfnamefont {Y.~Z.}\ \bibnamefont {{Xu}}}, \bibinfo {author} {\bibfnamefont {M.}~\bibnamefont {{Ding}}}, \bibinfo {author} {\bibfnamefont {K.}~\bibnamefont {{Raya}}}, \bibinfo {author} {\bibfnamefont {C.~D.}\ \bibnamefont {{Roberts}}}, \bibinfo {author} {\bibfnamefont {J.}~\bibnamefont {{Rodr{\'\i}guez-Quintero}}},\ and\ \bibinfo {author} {\bibfnamefont {S.~M.}\ \bibnamefont {{Schmidt}}},\ }\href {https://doi.org/10.1140/epjc/s10052-024-12518-x} {\bibfield  {journal} {\bibinfo  {journal} {European Physical Journal C}\ }\textbf {\bibinfo {volume} {84}},\ \bibinfo {eid} {191} (\bibinfo {year} {2024})},\ \Eprint {https://arxiv.org/abs/2311.14832} {arXiv:2311.14832 [hep-ph]} \BibitemShut {NoStop}%
\bibitem [{\citenamefont {Li}\ and\ \citenamefont {Vary}(2024)}]{YLi04}%
  \BibitemOpen
  \bibfield  {author} {\bibinfo {author} {\bibfnamefont {Y.}~\bibnamefont {Li}}\ and\ \bibinfo {author} {\bibfnamefont {J.~P.}\ \bibnamefont {Vary}},\ }\href {https://doi.org/10.1103/PhysRevD.109.L051501} {\bibfield  {journal} {\bibinfo  {journal} {Phys. Rev. D}\ }\textbf {\bibinfo {volume} {109}},\ \bibinfo {pages} {L051501} (\bibinfo {year} {2024})}\BibitemShut {NoStop}%
\bibitem [{\citenamefont {Liu}\ \emph {et~al.}(2024{\natexlab{a}})\citenamefont {Liu}, \citenamefont {Shuryak}, \citenamefont {Weiss},\ and\ \citenamefont {Zahed}}]{Zahed1}%
  \BibitemOpen
  \bibfield  {author} {\bibinfo {author} {\bibfnamefont {W.-Y.}\ \bibnamefont {Liu}}, \bibinfo {author} {\bibfnamefont {E.}~\bibnamefont {Shuryak}}, \bibinfo {author} {\bibfnamefont {C.}~\bibnamefont {Weiss}},\ and\ \bibinfo {author} {\bibfnamefont {I.}~\bibnamefont {Zahed}},\ }\href {https://doi.org/10.1103/PhysRevD.110.054021} {\bibfield  {journal} {\bibinfo  {journal} {Phys. Rev. D}\ }\textbf {\bibinfo {volume} {110}},\ \bibinfo {pages} {054021} (\bibinfo {year} {2024}{\natexlab{a}})}\BibitemShut {NoStop}%
\bibitem [{\citenamefont {Liu}\ \emph {et~al.}(2024{\natexlab{b}})\citenamefont {Liu}, \citenamefont {Shuryak},\ and\ \citenamefont {Zahed}}]{Zahed2}%
  \BibitemOpen
  \bibfield  {author} {\bibinfo {author} {\bibfnamefont {W.-Y.}\ \bibnamefont {Liu}}, \bibinfo {author} {\bibfnamefont {E.}~\bibnamefont {Shuryak}},\ and\ \bibinfo {author} {\bibfnamefont {I.}~\bibnamefont {Zahed}},\ }\href {https://doi.org/10.1103/PhysRevD.110.054022} {\bibfield  {journal} {\bibinfo  {journal} {Phys. Rev. D}\ }\textbf {\bibinfo {volume} {110}},\ \bibinfo {pages} {054022} (\bibinfo {year} {2024}{\natexlab{b}})}\BibitemShut {NoStop}%
\bibitem [{\citenamefont {Broniowski}\ and\ \citenamefont {{Ruiz Arriola}}(2024)}]{WB24}%
  \BibitemOpen
  \bibfield  {author} {\bibinfo {author} {\bibfnamefont {W.}~\bibnamefont {Broniowski}}\ and\ \bibinfo {author} {\bibfnamefont {E.}~\bibnamefont {{Ruiz Arriola}}},\ }\href {https://doi.org/https://doi.org/10.1016/j.physletb.2024.139138} {\bibfield  {journal} {\bibinfo  {journal} {Physics Letters B}\ }\textbf {\bibinfo {volume} {859}},\ \bibinfo {pages} {139138} (\bibinfo {year} {2024})}\BibitemShut {NoStop}%
\bibitem [{\citenamefont {Brodsky}\ \emph {et~al.}(2001)\citenamefont {Brodsky}, \citenamefont {Hwang}, \citenamefont {Ma},\ and\ \citenamefont {Schmidt}}]{BD2001}%
  \BibitemOpen
  \bibfield  {author} {\bibinfo {author} {\bibfnamefont {S.~J.}\ \bibnamefont {Brodsky}}, \bibinfo {author} {\bibfnamefont {D.~S.}\ \bibnamefont {Hwang}}, \bibinfo {author} {\bibfnamefont {B.-Q.}\ \bibnamefont {Ma}},\ and\ \bibinfo {author} {\bibfnamefont {I.}~\bibnamefont {Schmidt}},\ }\href {https://doi.org/https://doi.org/10.1016/S0550-3213(00)00626-X} {\bibfield  {journal} {\bibinfo  {journal} {Nuclear Physics B}\ }\textbf {\bibinfo {volume} {593}},\ \bibinfo {pages} {311} (\bibinfo {year} {2001})}\BibitemShut {NoStop}%
\bibitem [{\citenamefont {Choi}\ and\ \citenamefont {Ji}(2014)}]{CJ14}%
  \BibitemOpen
  \bibfield  {author} {\bibinfo {author} {\bibfnamefont {H.-M.}\ \bibnamefont {Choi}}\ and\ \bibinfo {author} {\bibfnamefont {C.-R.}\ \bibnamefont {Ji}},\ }\href {https://doi.org/10.1103/PhysRevD.89.033011} {\bibfield  {journal} {\bibinfo  {journal} {Phys. Rev. D}\ }\textbf {\bibinfo {volume} {89}},\ \bibinfo {pages} {033011} (\bibinfo {year} {2014})}\BibitemShut {NoStop}%
\bibitem [{\citenamefont {Choi}\ and\ \citenamefont {Ji}(2015)}]{CJ15}%
  \BibitemOpen
  \bibfield  {author} {\bibinfo {author} {\bibfnamefont {H.-M.}\ \bibnamefont {Choi}}\ and\ \bibinfo {author} {\bibfnamefont {C.-R.}\ \bibnamefont {Ji}},\ }\href {https://doi.org/10.1103/PhysRevD.91.014018} {\bibfield  {journal} {\bibinfo  {journal} {Phys. Rev. D}\ }\textbf {\bibinfo {volume} {91}},\ \bibinfo {pages} {014018} (\bibinfo {year} {2015})}\BibitemShut {NoStop}%
\bibitem [{\citenamefont {Choi}\ and\ \citenamefont {Ji}(2017)}]{CJ17}%
  \BibitemOpen
  \bibfield  {author} {\bibinfo {author} {\bibfnamefont {H.-M.}\ \bibnamefont {Choi}}\ and\ \bibinfo {author} {\bibfnamefont {C.-R.}\ \bibnamefont {Ji}},\ }\href {https://doi.org/10.1103/PhysRevD.95.056002} {\bibfield  {journal} {\bibinfo  {journal} {Phys. Rev. D}\ }\textbf {\bibinfo {volume} {95}},\ \bibinfo {pages} {056002} (\bibinfo {year} {2017})}\BibitemShut {NoStop}%
\bibitem [{\citenamefont {Arifi}\ \emph {et~al.}(2023{\natexlab{a}})\citenamefont {Arifi}, \citenamefont {Choi}, \citenamefont {Ji},\ and\ \citenamefont {Oh}}]{Jafar1}%
  \BibitemOpen
  \bibfield  {author} {\bibinfo {author} {\bibfnamefont {A.~J.}\ \bibnamefont {Arifi}}, \bibinfo {author} {\bibfnamefont {H.-M.}\ \bibnamefont {Choi}}, \bibinfo {author} {\bibfnamefont {C.-R.}\ \bibnamefont {Ji}},\ and\ \bibinfo {author} {\bibfnamefont {Y.}~\bibnamefont {Oh}},\ }\href {https://doi.org/10.1103/PhysRevD.107.053003} {\bibfield  {journal} {\bibinfo  {journal} {Phys. Rev. D}\ }\textbf {\bibinfo {volume} {107}},\ \bibinfo {pages} {053003} (\bibinfo {year} {2023}{\natexlab{a}})}\BibitemShut {NoStop}%
\bibitem [{\citenamefont {Arifi}\ \emph {et~al.}(2023{\natexlab{b}})\citenamefont {Arifi}, \citenamefont {Choi},\ and\ \citenamefont {Ji}}]{Jafar2}%
  \BibitemOpen
  \bibfield  {author} {\bibinfo {author} {\bibfnamefont {A.~J.}\ \bibnamefont {Arifi}}, \bibinfo {author} {\bibfnamefont {H.-M.}\ \bibnamefont {Choi}},\ and\ \bibinfo {author} {\bibfnamefont {C.-R.}\ \bibnamefont {Ji}},\ }\href {https://doi.org/10.1103/PhysRevD.108.013006} {\bibfield  {journal} {\bibinfo  {journal} {Phys. Rev. D}\ }\textbf {\bibinfo {volume} {108}},\ \bibinfo {pages} {013006} (\bibinfo {year} {2023}{\natexlab{b}})}\BibitemShut {NoStop}%
\bibitem [{\citenamefont {Choi}(2021{\natexlab{a}})}]{Choi21}%
  \BibitemOpen
  \bibfield  {author} {\bibinfo {author} {\bibfnamefont {H.-M.}\ \bibnamefont {Choi}},\ }\href {https://doi.org/10.1103/PhysRevD.103.073004} {\bibfield  {journal} {\bibinfo  {journal} {Phys. Rev. D}\ }\textbf {\bibinfo {volume} {103}},\ \bibinfo {pages} {073004} (\bibinfo {year} {2021}{\natexlab{a}})}\BibitemShut {NoStop}%
\bibitem [{\citenamefont {Choi}(2021{\natexlab{b}})}]{ChoiAdv}%
  \BibitemOpen
  \bibfield  {author} {\bibinfo {author} {\bibfnamefont {H.-M.}\ \bibnamefont {Choi}},\ }\href {https://doi.org/10.1155/2021/4277321} {\bibfield  {journal} {\bibinfo  {journal} {Adv. High Energy Phys.}\ }\textbf {\bibinfo {volume} {2021}},\ \bibinfo {pages} {4277321} (\bibinfo {year} {2021}{\natexlab{b}})},\ \Eprint {https://arxiv.org/abs/2108.10544} {arXiv:2108.10544 [hep-ph]} \BibitemShut {NoStop}%
\bibitem [{\citenamefont {Choi}\ and\ \citenamefont {Ji}(2024)}]{Choi:2024ptc}%
  \BibitemOpen
  \bibfield  {author} {\bibinfo {author} {\bibfnamefont {H.-M.}\ \bibnamefont {Choi}}\ and\ \bibinfo {author} {\bibfnamefont {C.-R.}\ \bibnamefont {Ji}},\ }\href {https://doi.org/10.1103/PhysRevD.110.014006} {\bibfield  {journal} {\bibinfo  {journal} {Phys. Rev. D}\ }\textbf {\bibinfo {volume} {110}},\ \bibinfo {pages} {014006} (\bibinfo {year} {2024})},\ \Eprint {https://arxiv.org/abs/2403.16703} {arXiv:2403.16703 [hep-ph]} \BibitemShut {NoStop}%
\bibitem [{\citenamefont {Bakamjian}\ and\ \citenamefont {Thomas}(1953)}]{BT53}%
  \BibitemOpen
  \bibfield  {author} {\bibinfo {author} {\bibfnamefont {B.}~\bibnamefont {Bakamjian}}\ and\ \bibinfo {author} {\bibfnamefont {L.~H.}\ \bibnamefont {Thomas}},\ }\href {https://doi.org/10.1103/PhysRev.92.1300} {\bibfield  {journal} {\bibinfo  {journal} {Phys. Rev.}\ }\textbf {\bibinfo {volume} {92}},\ \bibinfo {pages} {1300} (\bibinfo {year} {1953})}\BibitemShut {NoStop}%
\bibitem [{\citenamefont {Polyzou}(2010)}]{Poly10}%
  \BibitemOpen
  \bibfield  {author} {\bibinfo {author} {\bibfnamefont {W.~N.}\ \bibnamefont {Polyzou}},\ }\href {https://doi.org/10.1103/PhysRevC.82.064001} {\bibfield  {journal} {\bibinfo  {journal} {Phys. Rev. C}\ }\textbf {\bibinfo {volume} {82}},\ \bibinfo {pages} {064001} (\bibinfo {year} {2010})}\BibitemShut {NoStop}%
\bibitem [{\citenamefont {Chung}\ \emph {et~al.}(1988)\citenamefont {Chung}, \citenamefont {Coester}, \citenamefont {Keister},\ and\ \citenamefont {Polyzou}}]{CCKP}%
  \BibitemOpen
  \bibfield  {author} {\bibinfo {author} {\bibfnamefont {P.~L.}\ \bibnamefont {Chung}}, \bibinfo {author} {\bibfnamefont {F.}~\bibnamefont {Coester}}, \bibinfo {author} {\bibfnamefont {B.~D.}\ \bibnamefont {Keister}},\ and\ \bibinfo {author} {\bibfnamefont {W.~N.}\ \bibnamefont {Polyzou}},\ }\href {https://doi.org/10.1103/PhysRevC.37.2000} {\bibfield  {journal} {\bibinfo  {journal} {Phys. Rev. C}\ }\textbf {\bibinfo {volume} {37}},\ \bibinfo {pages} {2000} (\bibinfo {year} {1988})}\BibitemShut {NoStop}%
\bibitem [{\citenamefont {Choi}\ and\ \citenamefont {Ji}(1999{\natexlab{a}})}]{CJ97}%
  \BibitemOpen
  \bibfield  {author} {\bibinfo {author} {\bibfnamefont {H.-M.}\ \bibnamefont {Choi}}\ and\ \bibinfo {author} {\bibfnamefont {C.-R.}\ \bibnamefont {Ji}},\ }\href {https://doi.org/10.1103/PhysRevD.59.074015} {\bibfield  {journal} {\bibinfo  {journal} {Phys. Rev. D}\ }\textbf {\bibinfo {volume} {59}},\ \bibinfo {pages} {074015} (\bibinfo {year} {1999}{\natexlab{a}})}\BibitemShut {NoStop}%
\bibitem [{\citenamefont {Choi}\ and\ \citenamefont {Ji}(1999{\natexlab{b}})}]{CJ99a}%
  \BibitemOpen
  \bibfield  {author} {\bibinfo {author} {\bibfnamefont {H.-M.}\ \bibnamefont {Choi}}\ and\ \bibinfo {author} {\bibfnamefont {C.-R.}\ \bibnamefont {Ji}},\ }\href {https://doi.org/https://doi.org/10.1016/S0370-2693(99)00817-5} {\bibfield  {journal} {\bibinfo  {journal} {Physics Letters B}\ }\textbf {\bibinfo {volume} {460}},\ \bibinfo {pages} {461} (\bibinfo {year} {1999}{\natexlab{b}})}\BibitemShut {NoStop}%
\bibitem [{\citenamefont {Choi}\ \emph {et~al.}(2015)\citenamefont {Choi}, \citenamefont {Ji}, \citenamefont {Li},\ and\ \citenamefont {Ryu}}]{CJLR15}%
  \BibitemOpen
  \bibfield  {author} {\bibinfo {author} {\bibfnamefont {H.-M.}\ \bibnamefont {Choi}}, \bibinfo {author} {\bibfnamefont {C.-R.}\ \bibnamefont {Ji}}, \bibinfo {author} {\bibfnamefont {Z.}~\bibnamefont {Li}},\ and\ \bibinfo {author} {\bibfnamefont {H.-Y.}\ \bibnamefont {Ryu}},\ }\href {https://doi.org/10.1103/PhysRevC.92.055203} {\bibfield  {journal} {\bibinfo  {journal} {Phys. Rev. C}\ }\textbf {\bibinfo {volume} {92}},\ \bibinfo {pages} {055203} (\bibinfo {year} {2015})}\BibitemShut {NoStop}%
\bibitem [{\citenamefont {Arifi}\ \emph {et~al.}(2022)\citenamefont {Arifi}, \citenamefont {Choi}, \citenamefont {Ji},\ and\ \citenamefont {Oh}}]{ACJO22}%
  \BibitemOpen
  \bibfield  {author} {\bibinfo {author} {\bibfnamefont {A.~J.}\ \bibnamefont {Arifi}}, \bibinfo {author} {\bibfnamefont {H.-M.}\ \bibnamefont {Choi}}, \bibinfo {author} {\bibfnamefont {C.-R.}\ \bibnamefont {Ji}},\ and\ \bibinfo {author} {\bibfnamefont {Y.}~\bibnamefont {Oh}},\ }\href {https://doi.org/10.1103/PhysRevD.106.014009} {\bibfield  {journal} {\bibinfo  {journal} {Phys. Rev. D}\ }\textbf {\bibinfo {volume} {106}},\ \bibinfo {pages} {014009} (\bibinfo {year} {2022})}\BibitemShut {NoStop}%
\bibitem [{\citenamefont {Jaus}(1991)}]{SLF2}%
  \BibitemOpen
  \bibfield  {author} {\bibinfo {author} {\bibfnamefont {W.}~\bibnamefont {Jaus}},\ }\href {https://doi.org/10.1103/PhysRevD.44.2851} {\bibfield  {journal} {\bibinfo  {journal} {Phys. Rev. D}\ }\textbf {\bibinfo {volume} {44}},\ \bibinfo {pages} {2851} (\bibinfo {year} {1991})}\BibitemShut {NoStop}%
\bibitem [{\citenamefont {Jaus}(1990)}]{SLF3}%
  \BibitemOpen
  \bibfield  {author} {\bibinfo {author} {\bibfnamefont {W.}~\bibnamefont {Jaus}},\ }\href {https://doi.org/10.1103/PhysRevD.41.3394} {\bibfield  {journal} {\bibinfo  {journal} {Phys. Rev. D}\ }\textbf {\bibinfo {volume} {41}},\ \bibinfo {pages} {3394} (\bibinfo {year} {1990})}\BibitemShut {NoStop}%
\bibitem [{\citenamefont {Cheng}\ \emph {et~al.}(1997)\citenamefont {Cheng}, \citenamefont {Cheung},\ and\ \citenamefont {Hwang}}]{CCC97}%
  \BibitemOpen
  \bibfield  {author} {\bibinfo {author} {\bibfnamefont {H.-Y.}\ \bibnamefont {Cheng}}, \bibinfo {author} {\bibfnamefont {C.-Y.}\ \bibnamefont {Cheung}},\ and\ \bibinfo {author} {\bibfnamefont {C.-W.}\ \bibnamefont {Hwang}},\ }\href {https://doi.org/10.1103/PhysRevD.55.1559} {\bibfield  {journal} {\bibinfo  {journal} {Phys. Rev. D}\ }\textbf {\bibinfo {volume} {55}},\ \bibinfo {pages} {1559} (\bibinfo {year} {1997})},\ \Eprint {https://arxiv.org/abs/hep-ph/9607332} {arXiv:hep-ph/9607332} \BibitemShut {NoStop}%
\bibitem [{\citenamefont {Coester}\ and\ \citenamefont {Polyzou}(2005)}]{CP05}%
  \BibitemOpen
  \bibfield  {author} {\bibinfo {author} {\bibfnamefont {F.}~\bibnamefont {Coester}}\ and\ \bibinfo {author} {\bibfnamefont {W.~N.}\ \bibnamefont {Polyzou}},\ }\href {https://doi.org/10.1103/PhysRevC.71.028202} {\bibfield  {journal} {\bibinfo  {journal} {Phys. Rev. C}\ }\textbf {\bibinfo {volume} {71}},\ \bibinfo {pages} {028202} (\bibinfo {year} {2005})}\BibitemShut {NoStop}%
\bibitem [{\citenamefont {Choi}\ and\ \citenamefont {Ji}(2007)}]{CJ07}%
  \BibitemOpen
  \bibfield  {author} {\bibinfo {author} {\bibfnamefont {H.-M.}\ \bibnamefont {Choi}}\ and\ \bibinfo {author} {\bibfnamefont {C.-R.}\ \bibnamefont {Ji}},\ }\href {https://doi.org/10.1103/PhysRevD.75.034019} {\bibfield  {journal} {\bibinfo  {journal} {Phys. Rev. D}\ }\textbf {\bibinfo {volume} {75}},\ \bibinfo {pages} {034019} (\bibinfo {year} {2007})}\BibitemShut {NoStop}%
\bibitem [{\citenamefont {Melosh}(1974)}]{Melosh}%
  \BibitemOpen
  \bibfield  {author} {\bibinfo {author} {\bibfnamefont {H.~J.}\ \bibnamefont {Melosh}},\ }\href {https://doi.org/10.1103/PhysRevD.9.1095} {\bibfield  {journal} {\bibinfo  {journal} {Phys. Rev. D}\ }\textbf {\bibinfo {volume} {9}},\ \bibinfo {pages} {1095} (\bibinfo {year} {1974})}\BibitemShut {NoStop}%
\bibitem [{\citenamefont {Choi}\ and\ \citenamefont {Ji}(2009)}]{CJ09}%
  \BibitemOpen
  \bibfield  {author} {\bibinfo {author} {\bibfnamefont {H.-M.}\ \bibnamefont {Choi}}\ and\ \bibinfo {author} {\bibfnamefont {C.-R.}\ \bibnamefont {Ji}},\ }\href {https://doi.org/10.1103/PhysRevD.80.054016} {\bibfield  {journal} {\bibinfo  {journal} {Phys. Rev. D}\ }\textbf {\bibinfo {volume} {80}},\ \bibinfo {pages} {054016} (\bibinfo {year} {2009})}\BibitemShut {NoStop}%
\bibitem [{\citenamefont {Dhiman}\ \emph {et~al.}(2019)\citenamefont {Dhiman}, \citenamefont {Dahiya}, \citenamefont {Ji},\ and\ \citenamefont {Choi}}]{Nisha}%
  \BibitemOpen
  \bibfield  {author} {\bibinfo {author} {\bibfnamefont {N.}~\bibnamefont {Dhiman}}, \bibinfo {author} {\bibfnamefont {H.}~\bibnamefont {Dahiya}}, \bibinfo {author} {\bibfnamefont {C.-R.}\ \bibnamefont {Ji}},\ and\ \bibinfo {author} {\bibfnamefont {H.-M.}\ \bibnamefont {Choi}},\ }\href {https://doi.org/10.1103/PhysRevD.100.014026} {\bibfield  {journal} {\bibinfo  {journal} {Phys. Rev. D}\ }\textbf {\bibinfo {volume} {100}},\ \bibinfo {pages} {014026} (\bibinfo {year} {2019})}\BibitemShut {NoStop}%
\bibitem [{\citenamefont {Lepage}\ and\ \citenamefont {Brodsky}(1980)}]{BD80}%
  \BibitemOpen
  \bibfield  {author} {\bibinfo {author} {\bibfnamefont {G.~P.}\ \bibnamefont {Lepage}}\ and\ \bibinfo {author} {\bibfnamefont {S.~J.}\ \bibnamefont {Brodsky}},\ }\href {https://doi.org/10.1103/PhysRevD.22.2157} {\bibfield  {journal} {\bibinfo  {journal} {Phys. Rev. D}\ }\textbf {\bibinfo {volume} {22}},\ \bibinfo {pages} {2157} (\bibinfo {year} {1980})}\BibitemShut {NoStop}%
\bibitem [{\citenamefont {Navas}\ \emph {et~al.}(2024)\citenamefont {Navas} \emph {et~al.}}]{ParticleDataGroup:2024cfk}%
  \BibitemOpen
  \bibfield  {author} {\bibinfo {author} {\bibfnamefont {S.}~\bibnamefont {Navas}} \emph {et~al.} (\bibinfo {collaboration} {Particle Data Group}),\ }\href {https://doi.org/10.1103/PhysRevD.110.030001} {\bibfield  {journal} {\bibinfo  {journal} {Phys. Rev. D}\ }\textbf {\bibinfo {volume} {110}},\ \bibinfo {pages} {030001} (\bibinfo {year} {2024})}\BibitemShut {NoStop}%
\bibitem [{\citenamefont {{J. Volmer {\em et al.}}}(2001)}]{Volmer2001}%
  \BibitemOpen
  \bibfield  {author} {\bibinfo {author} {\bibnamefont {{J. Volmer {\em et al.}}}} (\bibinfo {collaboration} {The Jefferson Lab F\ensuremath{\pi} Collaboration}),\ }\href {https://doi.org/10.1103/PhysRevLett.86.1713} {\bibfield  {journal} {\bibinfo  {journal} {Phys. Rev. Lett.}\ }\textbf {\bibinfo {volume} {86}},\ \bibinfo {pages} {1713} (\bibinfo {year} {2001})}\BibitemShut {NoStop}%
\bibitem [{\citenamefont {{S.R. Amendolia {\em et al.}}}(1986)}]{Amendolia}%
  \BibitemOpen
  \bibfield  {author} {\bibinfo {author} {\bibnamefont {{S.R. Amendolia {\em et al.}}}},\ }\href {https://doi.org/https://doi.org/10.1016/0550-3213(86)90437-2} {\bibfield  {journal} {\bibinfo  {journal} {Nuclear Physics B}\ }\textbf {\bibinfo {volume} {277}},\ \bibinfo {pages} {168} (\bibinfo {year} {1986})}\BibitemShut {NoStop}%
\bibitem [{\citenamefont {{V. Tadevosyan {\em et al.}}}(2007)}]{Tad2007}%
  \BibitemOpen
  \bibfield  {author} {\bibinfo {author} {\bibnamefont {{V. Tadevosyan {\em et al.}}}},\ }\href {https://doi.org/10.1103/PhysRevC.75.055205} {\bibfield  {journal} {\bibinfo  {journal} {Phys. Rev. C}\ }\textbf {\bibinfo {volume} {75}},\ \bibinfo {pages} {055205} (\bibinfo {year} {2007})}\BibitemShut {NoStop}%
\bibitem [{\citenamefont {{T. Horn, {\em et al.} }}(2006)}]{Horn2006}%
  \BibitemOpen
  \bibfield  {author} {\bibinfo {author} {\bibnamefont {{T. Horn, {\em et al.} }}},\ }\href {https://doi.org/10.1103/PhysRevLett.97.192001} {\bibfield  {journal} {\bibinfo  {journal} {Phys. Rev. Lett.}\ }\textbf {\bibinfo {volume} {97}},\ \bibinfo {pages} {192001} (\bibinfo {year} {2006})}\BibitemShut {NoStop}%
\bibitem [{\citenamefont {Burkardt}(2003)}]{BURKARDT_2003}%
  \BibitemOpen
  \bibfield  {author} {\bibinfo {author} {\bibfnamefont {M.}~\bibnamefont {Burkardt}},\ }\href {https://doi.org/10.1142/s0217751x03012370} {\bibfield  {journal} {\bibinfo  {journal} {International Journal of Modern Physics A}\ }\textbf {\bibinfo {volume} {18}},\ \bibinfo {pages} {173–207} (\bibinfo {year} {2003})}\BibitemShut {NoStop}%
\bibitem [{\citenamefont {Freese}\ and\ \citenamefont {Miller}(2021)}]{Freese:2021czn}%
  \BibitemOpen
  \bibfield  {author} {\bibinfo {author} {\bibfnamefont {A.}~\bibnamefont {Freese}}\ and\ \bibinfo {author} {\bibfnamefont {G.~A.}\ \bibnamefont {Miller}},\ }\href {https://doi.org/10.1103/PhysRevD.103.094023} {\bibfield  {journal} {\bibinfo  {journal} {Phys. Rev. D}\ }\textbf {\bibinfo {volume} {103}},\ \bibinfo {pages} {094023} (\bibinfo {year} {2021})},\ \Eprint {https://arxiv.org/abs/2102.01683} {arXiv:2102.01683 [hep-ph]} \BibitemShut {NoStop}%
\bibitem [{\citenamefont {Freese}\ and\ \citenamefont {Miller}(2022)}]{Freese:2021mzg}%
  \BibitemOpen
  \bibfield  {author} {\bibinfo {author} {\bibfnamefont {A.}~\bibnamefont {Freese}}\ and\ \bibinfo {author} {\bibfnamefont {G.~A.}\ \bibnamefont {Miller}},\ }\href {https://doi.org/10.1103/PhysRevD.105.014003} {\bibfield  {journal} {\bibinfo  {journal} {Phys. Rev. D}\ }\textbf {\bibinfo {volume} {105}},\ \bibinfo {pages} {014003} (\bibinfo {year} {2022})},\ \Eprint {https://arxiv.org/abs/2108.03301} {arXiv:2108.03301 [hep-ph]} \BibitemShut {NoStop}%
\bibitem [{\citenamefont {Freese}\ and\ \citenamefont {Miller}(2023)}]{Freese:2022fat}%
  \BibitemOpen
  \bibfield  {author} {\bibinfo {author} {\bibfnamefont {A.}~\bibnamefont {Freese}}\ and\ \bibinfo {author} {\bibfnamefont {G.~A.}\ \bibnamefont {Miller}},\ }\href {https://doi.org/10.1103/PhysRevD.108.034008} {\bibfield  {journal} {\bibinfo  {journal} {Phys. Rev. D}\ }\textbf {\bibinfo {volume} {108}},\ \bibinfo {pages} {034008} (\bibinfo {year} {2023})},\ \Eprint {https://arxiv.org/abs/2210.03807} {arXiv:2210.03807 [hep-ph]} \BibitemShut {NoStop}%
\bibitem [{\citenamefont {Panteleeva}\ and\ \citenamefont {Polyakov}(2021)}]{Panteleeva:2021iip}%
  \BibitemOpen
  \bibfield  {author} {\bibinfo {author} {\bibfnamefont {J.~Y.}\ \bibnamefont {Panteleeva}}\ and\ \bibinfo {author} {\bibfnamefont {M.~V.}\ \bibnamefont {Polyakov}},\ }\href {https://doi.org/10.1103/PhysRevD.104.014008} {\bibfield  {journal} {\bibinfo  {journal} {Phys. Rev. D}\ }\textbf {\bibinfo {volume} {104}},\ \bibinfo {pages} {014008} (\bibinfo {year} {2021})},\ \Eprint {https://arxiv.org/abs/2102.10902} {arXiv:2102.10902 [hep-ph]} \BibitemShut {NoStop}%
\end{thebibliography}%
\end{document}